\documentclass[review]{elsarticle}
\usepackage{float}
\usepackage{verbatim} 
\usepackage{apalike}
\restylefloat{figure}
\floatstyle{plaintop} 
\restylefloat{table}

\usepackage{lineno}
\modulolinenumbers[5]
\usepackage{amsfonts}
\usepackage[colorlinks = true,
            linkcolor = blue,
            urlcolor  = blue,
            citecolor = blue,
            anchorcolor = blue]{hyperref}

\usepackage{enumitem}
\usepackage{graphicx}

\usepackage{caption}
\usepackage{subfig}
\usepackage{multirow}
\usepackage{multicol}
\usepackage{makecell}
\usepackage{longtable}
\usepackage{lipsum}
\usepackage{url}
\usepackage{booktabs}
\usepackage[table,xcdraw]{xcolor}
\usepackage{makecell}
\usepackage{mathtools}
\usepackage[left=2.5cm, right=2.5cm, top=2.5cm]{geometry}
\usepackage{algorithm}
\usepackage{algorithmic}
\newcolumntype{P}[1]{>{\centering\arraybackslash}p{#1}}

\usepackage{float}
\usepackage{multirow}
\usepackage{longtable}
\usepackage{tabulary}
\usepackage{array}
\usepackage{booktabs}
\usepackage[referable]{threeparttablex}
\usepackage{caption}

\makeatletter
\newcommand*\TY@cap@gobble[2][]{\\}
\def\ltabulary{%
    \def\caption{
        \@ifstar\TY@cap@gobble\TY@cap@gobble}
    \def\endfirsthead{\\}%
    \def\endhead{\\}%
    \def\endfoot{\\}%
    \def\endlastfoot{\\}%
    \def\tabulary{%
        \def\TY@final{%
    \def\endfirsthead{\LT@end@hd@ft\LT@firsthead}%
    \def\endhead{\LT@end@hd@ft\LT@head}%
    \def\endfoot{\LT@end@hd@ft\LT@foot}%
    \def\endlastfoot{\LT@end@hd@ft\LT@lastfoot}%
    \longtable}%
        \let\endTY@final\endlongtable
        \TY@tabular}%
    \dimen@\columnwidth
    \advance\dimen@-\LTleft
    \advance\dimen@-\LTright
    \tabulary\dimen@}

\makeatother
\newcolumntype{L}[1]{>{\raggedright\let\newline\\\arraybackslash\hspace{0pt}}m{#1}}
\newcolumntype{C}[1]{>{\centering\let\newline\\\arraybackslash\hspace{0pt}}m{#1}}
\newcolumntype{R}[1]{>{\raggedleft\let\newline\\\arraybackslash\hspace{0pt}}m{#1}}

\journal{Computers \& Chemical Engineering}

\biboptions{authoryear}

\begin{document}
\begin{frontmatter}











\title{Learning From Limited Data and Feedback for Cell Culture Process Monitoring: A Comparative Study}

\author[uts]{Johnny Peng \corref{cor1}}
\ead{johnny.peng@student.uts.edu.au}

\author[uts]{Thanh Tung Khuat}
\ead{thanhtung.khuat@uts.edu.au}

\author[csl]{Ellen Otte}
\ead{ellen.otte@csl.com.au}

\author[uts]{Katarzyna Musial}
\ead{Katarzyna.Musial-Gabrys@uts.edu.au}

\author[uts]{Bogdan Gabrys}
\ead{bogdan.gabrys@uts.edu.au}

\cortext[cor1]{Corresponding author.}
\address[uts]{Complex Adaptive Systems Laboratory, The Data Science Institute, University of Technology Sydney, NSW 2007, Australia}
\address[csl]{CSL Innovation, Melbourne, VIC 3000, Australia}

\begin{abstract}
In cell culture bioprocessing, real-time batch process monitoring (BPM) refers to the continuous tracking and analysis of key process variables—such as viable cell density, nutrient levels, metabolite concentrations, and product titer—throughout the duration of a batch run. This enables early detection of deviations and supports timely control actions to ensure optimal cell growth and product quality. BPM plays a critical role in ensuring the quality and regulatory compliance of biopharmaceutical manufacturing processes. However, the development of accurate soft sensors for BPM is hindered by key challenges, including limited historical data, infrequent feedback, heterogeneous process conditions, and high-dimensional sensory inputs. This study presents a comprehensive benchmarking analysis of machine learning (ML) methods designed to address these challenges, with a focus on learning from historical data with limited volume and relevance in the context of bioprocess monitoring. We evaluate multiple ML approaches—including feature dimensionality reduction, online learning, and just-in-time learning—across three datasets, one in silico dataset and two real-world experimental datasets. Our findings highlight the importance of training strategies in handling limited data and feedback, with batch learning proving effective in homogeneous settings, while just-in-time learning and online learning demonstrate superior adaptability in cold-start scenarios. Additionally, we identify key meta-features, such as feed media composition and process control strategies, that significantly impact model transferability. The results also suggest that integrating Raman-based predictions with lagged offline measurements enhances monitoring accuracy, offering a promising direction for future bioprocess soft sensor development.
\end{abstract}

\end{frontmatter}



\section{Introduction}
A commercial-scale biopharmaceutical manufacturing process typically operates in batch mode, encompassing key unit operations such as mammalian cell culture, downstream purification, and product formulation \citep{khba24}. To ensure product quality and regulatory compliance, real-time batch process monitoring (BPM) has emerged as a valuable tool for continuous process verification of key performance indicators (KPIs) such as titer, productivity, and other critical quality attributes (CQAs) \citep{khba24,UNDEY2009177,Jiang2017-kl}. Raman spectroscopy, a non-destructive analytical technique for molecular characterisation, is increasingly applied in BPM to monitor bioprocesses in real time \citep{escu2017}. It captures spectral signatures that reflect culture environments influenced by key and critical process parameters (KPPs and CPPs), including media composition and bioreactor conditions, enabling timely detection of abnormal patterns and deviations in CQAs. 

Despite having great potential, building soft sensors for BPM faces many difficulties. The first common challenge is the \textbf{limited availability of training data}. Small data issues are common in biopharmaceutical manufacturing operations because new biopharmaceutical products often have a short production history with limited bioreactor runs and batches at a manufacturing facility \citep{Brunner2021-sp}. Furthermore, real-time BPM soft sensors are primarily developed to monitor difficult-to-measure CQAs and CPPs, as they could benefit the most from having automated real-time monitoring by soft sensors. However, since these variables are difficult to obtain, \textbf{feedback on them is generally limited throughout the bioreactor run}. e.g. once or twice per day. This not only means that it is difficult to collect additional training data, but it also makes it challenging for the user to verify and be confident in the accuracy of the real-time soft sensors during operation. At the same time, with limited feedback, soft sensors cannot correct themselves promptly and produce misleading predictions between feedbacks.

Additionally, the cell lines, media compositions, and other experimental settings could vary from run to run, which can significantly impact the Raman pattern of the bioprocess, causing \textbf{heterogeneous data distributions across runs}, and severely compromise the transferability of the learning between runs \citep{Weiss2016}. This is commonly referred to as a "cold start" problem, which refers to the difficulties of making accurate predictions or recommendations when there is insufficient historical data with distribution specific to the current context or configuration \citep{Frri11}. The cold start scenario is commonly found during the development and optimisation phase of the bioprocess. During this phase, the Design of Experiment (DoE) approach would generally be taken, which introduces differences in configuration between runs, such as changing feed and fill media composition, allowing scientists to understand and optimise the process further. Conversely, a ``warm start" scenario would occur if there were existing data from bioreactors run with the same or similar configurations. In this case, the availability of such configuration-specific data mitigates the cold start problem, enabling the bioreactor monitoring system to monitor the KPPs and CPPs throughout the runs accurately. Warm start scenarios typically happen for production-scale bioreactor runs as a fixed range of optimal settings would be used all the time. However, since warm-start scenarios are generally a lot easier to tackle in practice, this benchmark study focuses on the \textit{cold-start} problem.

The last challenge arises from the input feature of the real-time monitoring model, the Raman Spectrum, which typically comprises thousands of wavenumbers. In contrast, the target variables, such as the offline measurements, were only captured once or twice per day, resulting in approximately 14-28 data points over a two-week bioprocess run. Thus, even if we have historical data accumulated from multiple bioreactor runs in the past, \textbf{the number of training data points is still relatively small compared to the feature dimensionality}, which causes convergence and overfitting issues for most of the ML models. 

With limited historical data and feedback, as well as heterogeneous data and high-dimensional input features, it is challenging to construct high-performing predictive models for difficult and expensive-to-measure variables of interest (KPPs and CPPs) in upstream processes. These challenges, and many more, had already been discussed in a series of publications concerned with adaptive soft sensors in the broader process industry over a decade ago (\cite{kaga09,kaga09a,kagr11}) and while a number of partial solutions have been proposed (\cite{kaga09b,kaga09c,kaga09d,kaga10a,kaga10,sabu16a,baga17}), many of the chalenges still remain, particularly in the BPM context. To address these fundamental challenges, this benchmarking exercise aims to explore and critically assess the effectiveness of various state-of-the-art ML methods in addressing these issues for bioprocess real-time monitoring tasks. Through designed experiments, these methods were evaluated on datasets containing runs with different characteristics, offering insights into the optimal strategies for building an accurate and robust model for real-time bioprocess monitoring of difficult or expensive-to-measure variables under cold-start scenarios with limited data. 
 
\section{Research Problem and Aims}
We previously conducted a literature review and identified a set of ML methods that could effectively address the challenges of limited data and feedback discussed in the previous section. For this benchmark study, we focus on three of these types of methods -- Dimensionality Reduction (DR), Just-In-Time Learning (JITL), and Online Learning (OL) -- with a summary of their respective strengths and corresponding research challenges presented in Table~\ref{tab:research_problem}. DR is particularly useful when the number of features is high relative to the available data, as it reduces dimensionality and mitigates overfitting while preserving essential information. JITL, on the other hand, is designed to cope with cold-start problems by selecting relevant training instances close to the target context, enabling better generalisation in heterogeneous data environments. On the other hand, OL attempts to address the cold start problem by enabling fast and continual model adaptation as new data arrives, which is crucial in settings with infrequent feedback and shifting data distributions. Notably, JITL appears especially promising as it addresses both limited data homogeneity and infrequent feedback. While all three methods have shown promise across various domains, their comparative effectiveness in the context of BPM for upstream bioprocessing remains underexplored. This study, therefore, benchmarks their performance in this specific application setting using a series of carefully designed experiments.

The remainder of this paper is organised as follows. Section~\ref{sec:Design} presents the benchmark experiment design, detailing the datasets used in this study, including the one in silico dataset and two experimental datasets. This section also outlines the overall experimental pipeline design and details about individual components and methods. Section~\ref{sec:results} begins with a detailed analysis of the experimental results, focusing on the challenges posed by limited data availability. This includes a transferability analysis and a meta-analysis to examine the impact of meta-features on data distribution, highlighting the issue of limited data homogeneity across batches. Following this, a comparative evaluation of different training strategies, including OL and JITL, is presented to assess their effectiveness in addressing these challenges, including evaluation of results from an independent CSL bioreactor run, which allows us to analyse the effectiveness of the proposed approaches in handling cold-start scenarios. Finally, Section~\ref{sec:conclusion} concludes the study by summarising key insights and outlining future research directions in bioprocess soft sensor development.

\renewcommand{\arraystretch}{2}

\begin{table}[H]
\centering
\caption{Summary of research problems, their challenges, and ML methods for addressing them}
\label{tab:research_problem}
\begin{tabular}{@{}>{\centering\arraybackslash}p{5cm}>{\centering\arraybackslash}p{2cm}>{\centering\arraybackslash}p{8cm}@{}}
\toprule
 \textbf{Challenges}                    & \textbf{ML Methods} & \textbf{Why it could be effective}\\ 
\midrule
 \textbf{Limited data volume} relative to feature dimensionality causes model overfitting and inability to converge          & Dimensionality Reduction (DR)& DR compresses the input features to a much smaller dimension with minimal information loss\\
\midrule
\textbf{Limited data homogeneity} causing cold-start problem & Just-In-Time-Learning (JITL) & Curating training data based on their relevance to the inferencing data, which minimises the domain gaps in the training dataset and improves the generalisation of the model.\\                                   
\midrule                                   
 \textbf{Infrequent feedback} in a changing environment leads to slow adaptations to changes in data distribution & Online Learning (OL)                          & OL model learns incrementally as new information is received, allowing the model to adjust its parameters with each new data point and quickly adapt to changes in data.\\
 & Just-In-Time-Learning (JITL) & JITL makes ML models only learn from data points similar to the data point that is trying to predict, automatically adapt models to changes in data distribution\\
\bottomrule
\end{tabular}
\end{table}

\section{Benchmark Experiment Design}
\label{sec:Design}
\subsection{Datasets}
Three datasets were selected to systematically evaluate machine learning approaches for bioprocess monitoring under limited data constraints, each representing different real-world conditions and challenges. The IndPenSim in silico dataset \citep{GOLDRICK201570} is a simulated benchmark dataset that provides controlled experimental conditions and a large volume of synthetic bioprocess data, making it ideal for testing model transferability and performance in a structured environment. The CSL dataset, derived from real bioreactor runs, captures the complexities of industrial-scale cell culture processes, including variations in feed media and process conditions, offering a realistic but challenging small-data scenario. Lastly, the AstraZeneca dataset \citep{Ga2021} comprises historical bioprocess data with time-dependent offline measurements, allowing us to explore predictive modelling using lagged observations rather than Raman spectroscopy. The combination of these datasets ensures a comprehensive evaluation of different learning strategies, addressing both cold-start challenges and real-time monitoring under diverse data conditions.

\subsubsection{IndPenSim In Silico Dataset}

The in silico dataset was generated using an advanced mathematical simulation of a 100,000-litre penicillin fermentation system referenced as IndPenSim \citep{GOLDRICK201570}. The IndPenSim is a detailed simulation of an industrial-scale fed-batch fermentation process used for penicillin production. The primary objective of the simulation is to provide a benchmark for process systems analysis and control studies. It was developed and validated using historical data from a 100,000 L bioreactor industrial Penicillium production process, ensuring the simulation accurately reflects real industrial conditions. IndPenSim was developed using a mechanistic model incorporating various factors influencing the fermentation process, including key environmental effects such as dissolved oxygen, viscosity, temperature, pH, and dissolved carbon dioxide. It also considers the impacts of nitrogen and phenylacetic acid concentrations on the biomass and penicillin production rates. For convenience, 100 batches of already simulated data can be downloaded from the IndPenSim website \citep{goldrick_2019}, which will be used as the In Silico Dataset for our study. It consists of data from simulations with different control strategies, including:

\begin{itemize}
    \item Batches 1-30: Controlled by recipe-driven approach
    \item Batches 31-60: Controlled by operators
    \item Batches 61-90: Controlled by an Advanced Process Control (APC) using the Raman spectroscopy
    \item Batches 91-100: Contain faults resulting in process deviations.
\end{itemize}

For the IndPenSim dataset, we will use the simulated Raman spectrum data and penicillin concentration to train Raman-based ML models to perform real-time BPM tasks with different ML methods, this allows us to evaluate the effectiveness of those methods for addressing limited data and feedback problems (Table~\ref{tab:research_problem}) under different control strategies.

\subsubsection{CSL Experimental Dataset}

Our industry partner, CSL, provided the experiment dataset generated from 34 bioreactor runs, where 5-L bioreactors were used for all runs. Each run slightly differs in cell line, feed, and fill media composition. To validate the performance of ML models under a cold-start scenario, CSL also provided an additional batch consisting of 4 bioreactor runs, with two of the four runs using completely new cell lines, feed media and feeding strategies that were not used in the 34 runs. Performing real-time monitoring on these two runs is a great example of a cold-start problem. Thus, we have purposely omitted it from the main benchmark experimental runs and used it to verify the findings from the experimental results of the benchmark experiments. 

Nine offline measures are sampled during the bioreactor runs, including \textbf{Glucose, Glutamine, Glutamate, Lactate, Ammonia, Sodium, Potassium, Calcium}, and \textbf{Osmolality}. These offline measurements require manual effort to gather and generate, so they are generally available at a frequency of once to twice per day and are subject to human error. Each bioreactor run typically takes about two weeks, generating around 14-28 data points. Furthermore, apart from the offline measurements, we have also captured the Raman spectrum of the cell culture medium throughout the bioreactor run. Raman data has 3,325 dimensions (wavenumber range from 100 to 3425 cm$^{-1}$), which will be used to predict all nine different offline measures. Raman data are captured automatically and are generally available every 45 minutes unless manual interventions exist. 
Furthermore, because Raman spectra and offline measurements cannot be collected simultaneously, the timestamps of online Raman spectra and offline measurements always differ. Therefore, we have taken the necessary steps to map Raman spectra to their best corresponding offline measurements for constructing training datasets for machine learning models.

\subsubsection{AstraZeneca Experimental Dataset}

Another public dataset used in this study is from \cite{Ga2021}, which comprises data extracted from AstraZeneca's upstream process development and production databases. It includes Chinese Hamster Ovary (CHO) cell lines used in producing various antibody products collected over seven years (2010-2016). The dataset spans 106 cultures across different operational scales, from bench-top (5L) to manufacturing (500L) volumes. Each culture has at least 25 parameters recorded offline over up to 17 days. These parameters encompass Culture Days, Elapsed Culture Time (ECT), Viable Cell Density (VCD), Total Cell Density (TCD), Average Cell Compactness (ACC), Average Cell Diameter (ACD), pH, Cell Viability, Elapsed Generation Number (EGN), Average Cell Volume (ACV), Osmolality, Cumulative Population Doubling Level (CPDL), concentrations of Glutamine, Glutamate, Lactate, Ammonium, Glucose, Sodium, Potassium, and Bicarbonate, as well as Temperature, pCO$_2$, pO$_2$, Monomer Content of the Final Product, and Product Concentration ([mAb]).

The AstraZeneca dataset is a multivariate time series characterised by rank deficiency and comprising both time-dependent and time-independent parameters. It includes both the original data with missing values and an imputed version, with the latter being selected for this benchmark study. Notably, Raman spectroscopy data are not included in this dataset, precluding the possibility of real-time monitoring. Instead, we employ lagged offline measurement values and time-related features to perform one-step-ahead predictions for various offline analytes, including glutamine, glutamate, lactate, ammonium, glucose, sodium, potassium, bicarbonate, and product concentration ([mAb]). While one-step-ahead (one-day-ahead) prediction differs from the development of Raman-based soft sensors for real-time monitoring, it provides insights into the information contained within lagged observations, which may complement Raman data. Specifically, whereas Raman spectroscopy captures only a snapshot of the system at a given moment and cannot account for recent temporal trends, these trends can be inferred from lagged observations. Therefore, if models trained solely on lagged observations can accurately predict one-step-ahead offline measurements, integrating them with Raman data has the potential to enhance the accuracy and robustness of soft sensor models.

\subsection{Experimentation Pipeline and Methodology}
\label{sec:ExpDesign}
\begin{figure}[H]
  \centering
  \includegraphics[width=1\linewidth]{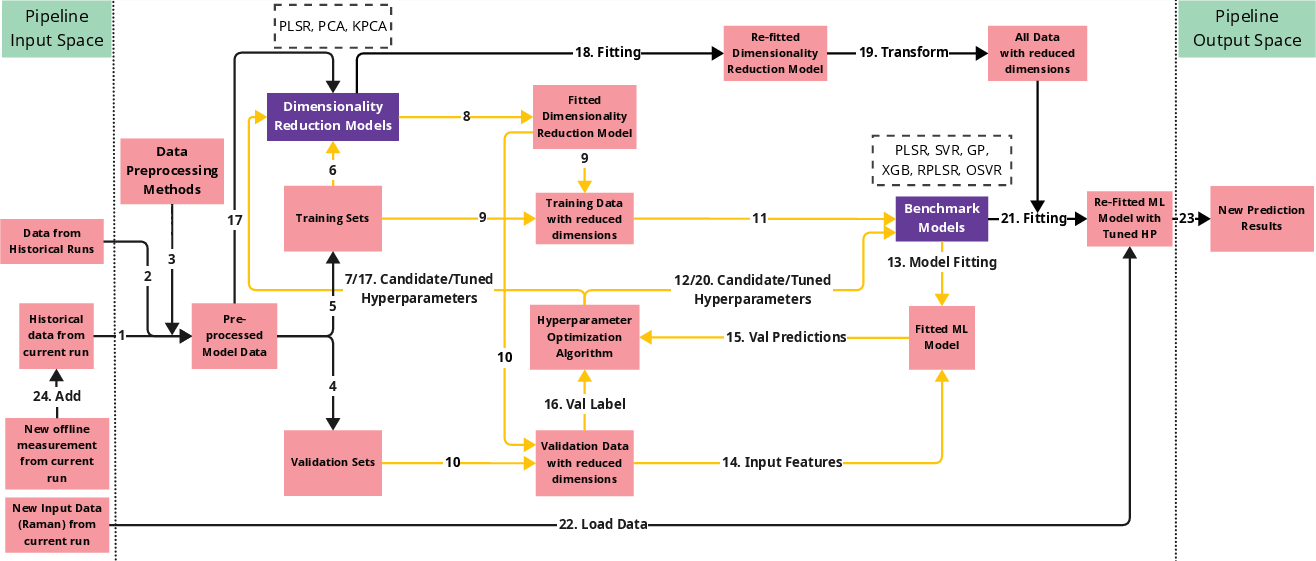}
  \caption{Experiment Pipeline Design}
  \label{Fig:BenchmarkExperimentDesign}
\end{figure}

Fig~\ref{Fig:BenchmarkExperimentDesign} shows the experiment pipeline we designed for training and evaluating the performance and effectiveness of each combination of DR techniques and ML models, which contains the following components:
\begin{itemize}
    \item Designate Benchmarking Components (purple blocks) - These components act as a standardised interface where different candidate methods (enumerated within the dotted-line boxes) can be placed in and systematically evaluated by running the pipeline. 
    \item Data States and Model Objects (pink blocks) - These blocks capture the evolving state of both data artifacts and trained model instances as they progress through various processing stages. 
    \item Input Output Space Marker (green blocks \& vertical dotted lines) - These objects visually bisect the pipeline and serve as conceptual separators between the input and output spaces.
    \item Hyperparameter-Optimisation (HPO) processes (yellow lines) - These steps are carried out \textbf{iteratively} until a set of optimised/tuned hyperparameters is generated. 
\end{itemize}

Furthermore, the ordering of the steps is numbered and can be grouped in stages shown below:
\begin{itemize}
    \item Data Preprocessing (1-3): Load and preprocess input data to the experiment pipeline by applying Trimming, SG Filter, and Normalization, in that order.
    \item Data Preparation (4 \& 5): Split data into a 5-fold training set and a validation set 
    \item HPO (6-16): The HPO process is used to find the optimal HPs for the prediction and DR models. 
    \item Training (17-20): Refit the DR model on all historical data using optimised HP, then transform all historical data into a lower dimension to refit the prediction model
    \item Inference and Update Database with New data (21-24): Generate new prediction/monitoring results on new Raman data and add new offline measurement data to the historical dataset when it becomes available 
\end{itemize}
Based on the designed experimental scenarios, experiments will be performed using this ML pipeline with different training strategies for predicting various offline measures with data from different bioprocess runs. Furthermore, multiple metrics are used to evaluate the performance of each model, including Normalised Mean Absolute Error (NMAE), Mean Absolute Percentage Error (MAPE), $R^2$, Root Mean Squared Error (RMSE), and Average Rank (Ranked by MAPE). For each offline measure, all performance metrics will be evaluated based on leave-one-batch-out cross-validation and averaged to compare them across different models and combinations. Lastly, following the recommended approach for comparing multiple model performances across multiple datasets \citep{demsar06a}, statistical testing was conducted using Friedman and Nemenyi's post hoc tests, and the results were visualised with critical difference plots. The average ranking used for statistical testing was computed based on NMAE, and it was chosen over MAPE because MAPE exhibits significant variance when the offline measurement is close to zero, which can occur during the early stages of the bioprocess run. Furthermore, NMAE is normalised against the difference between maximum and minimum values, which provides a more consistent comparison between models and bioprocess runs. More details about the individual components of this pipeline will be discussed in the following subsections.

\subsubsection{Data Preprocessing}

The only input features for CSL and IndPenSim data are the Raman spectra, which are noisy with many dimensions and naturally unsuitable for training ML models. To remove noisy signals from the data, a set of preprocessing methods was applied to improve its signal-to-noise ratio, thereby enhancing the performance of the downstream ML models, as demonstrated in an extensive study on preprocessing methods for Raman data by \cite{Poth2022}. For our research, Raman data are trimmed by removing wavenumbers in the $<500$ and $>3000$ regions. Then, the trimmed Raman data will be smoothed using the Savitzky–Golay filter, with a Polynomial order of 1 and a derivative order of 1. The window length was set to 25 for the CSL dataset and 15 for the IndPenSim dataset. Lastly, standard normal variate (SNV) normalisation is applied to the smoothed Raman data. After preprocessing is complete, the preprocessed data will be passed through a dimensionality reduction method to reduce the input dimensions to a number suitable for the ML model to fit, and more details about these methods will be discussed in later sections. 

For the AstraZeneca dataset, since Raman data is not available, the primary input features are time (in days since the run started) and lagged observations of all offline measures. One limitation of modelling on lagged observations is that features with more lags become available as the run progresses. For example, glucose measurements from three days ago will not become available until day 3. However, we cannot remove those yet-to-be-available observations from the feature set, as all training data points must have the same input dimension. Thus, we imputed those lagged observations as zero until they became available. Such an imputation is intuitive because it is equal to measuring an empty bioreactor before the run starts. However, the disadvantage of this approach is that it would result in highly sparse input features with many dimensions. Thus, similar to the CSL and IndPenSim datasets, dimensionality reduction methods will be applied to the preprocessed data.

\subsubsection{Machine Learning Models}
This study benchmarks a range of ML models for building Raman-based soft sensors and one-step-ahead offline measurement prediction, including:

\begin{itemize}
    \item Commonly Used Methods — SVR and PLSR. Based on the results from our literature review\cite{petu25}, PLSR and SVR are two of the most common methods used by industry and research for building Raman-based soft sensors. Thus, their performance could be a great reference point for understanding the performance of other ML models.
    \item Bayesian Method — GP. As discussed in our literature review, Bayesian ML methods have shown promising results for small data scenarios. Thus, we have also included the GP in our experiment.
    \item Tree-Based Method — XGBoost. Tree-based methods have shown promising results for structured data across various domains and great results for industry applications and Kaggle competitions. Thus, we have included the most advanced and commonly used tree model, XGBoost.
    \item Online Learning Models — Recursive PLSR (RPLSR) and Online SVR (OSVR). Our literature review discussed a few advantages of OL models when learning from limited data. In particlar, OL models can adapts to new data without retraining and hence does not subject to the risk of catastrophic forgetting. Furthmore, OL can be adjusted to put more weights on new data, which allows them to adapt quicker to changes in data. Thus, we have included the OL variation of PLSR and SVR, the two commonly used methods in academic research and industry for building Raman-based soft sensors. 
\end{itemize}

\subsubsection{Dimensionality Reduction Method}

Each raw Raman data typically has over 3,000 dimensions representing the wavenumber shifts due to the energy loss (gain) from scattering over the measuring material. Even after trimming, Raman spectrum data will still have close to 3,000 dimensions. Thus, it is common practice to perform DR on it before using it for model training. PCA-based methods have always been the main choices in the ML community for DR when building predictive ML models. Thus, we will be applying two types of PCA-based DR methods: Principal Component Analysis (PCA) and Kernel Principal Component Analysis (KPCA) to reduce the dimension of the input data. In addition, PLSR will also be used as a supervised DR method, which may perform better than the two PCA-based unsupervised DR methods as supervised information may help DR methods prioritise information that that are important for the ML task. Lastly, it is important to note that the PLSR and RPLSR models already perform DR as part of the modelling process. Thus, DR will not be applied to the Raman data before fitting these two models. 

\subsubsection{Hyperparameter Tuning and Testing Strategy}

Hyperparameter selection plays a critical role in determining the performance of machine learning models and can significantly impact experimental outcomes. In this benchmark study, hyperparameter tuning consists of two main steps. First, the train-validation split is performed to ensure robust model evaluation. Given the inherent phase-dependent variations in bioprocessing data, where metabolic and growth rates fluctuate across different phases, we employ leave-one-batch-out cross-validation. In this approach, one batch is held out as the test set, while the remaining batches serve as training data. This process is repeated across all batches to ensure phase-independent validation. Second, a hyperparameter search is conducted for machine learning and DR models to find a hyperparameter set that holistically optimises the end-to-end ML pipeline. We utilise the Tree-structured Parzen Estimator algorithm from the Python package Optuna \citep{tash2019} to perform 100 trials with a 5-fold train-validation split, optimising hyperparameters to minimise the average RMSE across all folds.

\subsubsection{Training Strategies}
To evaluate the effectiveness of different model training approaches for real-time monitoring in bioprocessing, we compare four distinct training strategies: \textbf{Pre-training, Retraining, Just-in-Time Learning (JITL), and Online Learning (OL)}. These strategies differ in how and when model updates occur, affecting their adaptability and computational efficiency. Below, we outline the key characteristics and procedures associated with each approach.

\paragraph{Training Strategy 0 – No Training (Baseline Models)}\hfill\\  
\textbf{Retraining Trigger:} None.  
The no-training approach utilises baseline models that require no feature inputs or machine learning training. These models serve as a lower bound for predictive performance, ensuring that any trained machine learning model should outperform them. The baseline methods included in this benchmark are:
    \begin{itemize}
        \item Last Value (LV): Predicts future offline measurements using the last recorded offline value until a new measurement is available.
        \item Mean of the Last Two Values (MLTV): Uses the average of the last two recorded offline measurements as the predicted future value.     
        \item Extrapolate of the Last Two Values (ELTV): Uses the linear extrapolated value of the last two recorded offline measurements as the predicted future value.     
    \end{itemize}
Since these models rely solely on past offline measurements without considering Raman spectra or any other features, their predictions represent the minimal achievable performance. Despite their simplicity, baseline models are useful for evaluating the added value of machine learning approaches by providing a reference point against which trained models are compared.

\paragraph{Training Strategy 1 – Pre-training}\hfill\\  
\textbf{Retraining Trigger:} None.  
In the \textbf{pre-training} approach, the model is trained once before the bioreactor run begins and remains unchanged throughout the process. This strategy relies solely on historical data to develop a predictive model, which is then applied without further updates. The steps involved are:  
\begin{enumerate}
    \item Split all historical data into training and validation sets.
    \item Perform hyperparameter optimisation (HPO) and determine an optimal set of hyperparameters.
    \item Train a machine learning model on the entire historical dataset using the tuned hyperparameters. The trained model is then used to generate real-time predictions for offline measurements without any further updates during the bioreactor run.
\end{enumerate}

\paragraph{Training Strategy 2 – Retraining}\hfill\\
\textbf{Retraining Trigger:} Whenever new offline data becomes available.  
The \textbf{retraining} approach extends pre-training by allowing model updates at fixed intervals or whenever new offline measurements arrive. This strategy enables adaptability while maintaining computational efficiency. The retraining steps are as follows:  
\begin{enumerate}
    \item Train the initial machine learning model using all historical data and the optimised hyperparameters from pre-training.
    \item When new offline measurements become available, retrain the model using the combined historical and newly acquired data while keeping the pre-determined hyperparameters fixed.
    \item The updated model generates real-time monitoring results based on Raman spectroscopy data until the next offline measurement is available.
\end{enumerate}

\paragraph{Training Strategy 3 – Just-in-Time Learning (JITL)}\hfill\\
\textbf{Retraining Trigger:} At every inference step.  
The \textbf{JITL} strategy retrains the model dynamically at each inference step using only a subset of relevant historical data points. This ensures that predictions are always based on the most relevant prior observations. The steps include:  
\begin{enumerate}
    \item Identify the \textbf{k = 30} most similar historical data points to the current Raman spectroscopy input, referred to as the "selected data". K = 30 was chosen to replicate the implementation from study \citet{Tulsyan2021} on JITL for similar use cases, which has shown great improvement over batch learning global models.
    \item Using the hyperparameters obtained during pre-training, refit both the dimensionality reduction and predictive models to the selected data.
    \item Use the newly trained model to generate real-time predictions, after which the model is discarded.
\end{enumerate}

\paragraph{Training Strategy 4 – Online Learning (OL)}\hfill\\ 
\textbf{Retraining Trigger:} Whenever new offline data becomes available.  
The OL approach enables continuous model adaptation by incrementally updating the model parameters as new offline measurements become available. This allows the model to dynamically adjust to changes in the bioprocess without requiring full retraining. The steps include:  
\begin{enumerate}
    \item Train an initial OL model using historical data and optimised hyperparameters from pre-training.
    \item Use the trained model to generate real-time monitoring results based on new Raman spectroscopy data.
    \item Incrementally update the model whenever new offline data becomes available, ensuring that it continuously learns from the most recent observations.
\end{enumerate}

By comparing these four strategies, we aim to assess their respective advantages and limitations in terms of prediction accuracy, computational efficiency, and adaptability to changing bioprocess conditions.

\section{Experiment and Results}
\label{sec:results}
\subsection{Transferability Analysis}
To highlight the prevalence of the data heterogeneity issue and the severity of the cold-start problem, we have performed a \textbf{systematic transferability analysis} between all historical batches, where we train a model on one batch of data (training batch) and evaluate its performance for real-time monitoring of another batch (target batch). This experiment was then repeatedly carried out for all possible pair-wise combinations of batches. Furthermore, since we are using only one batch of data to train the ML models at a time, e.g. less than 40 data points, it allows us to explore the effectiveness of each DR and ML model under a limited data scenario. For comparing the performance of different methods, we followed the recommended approach for comparing multiple model performance over multiple datasets \cite{demsar06a}, and statistical testing was done using Friedman and Nemenyi's posthoc tests and visualised with a Critical Difference Plot (CDP). 

Based on the results from this transferability analysis, we found two main factors that could impact models' transferability - \textbf{the homogeneity between the training batch and target batch} and \textbf{the complexity of dimensionality reduction and prediction models}. A more detailed analysis of why and how these factors impact the model's transferability is discussed in the following sections.  

\subsubsection{CSL datasets}
\begin{figure}[H]
    \centering
    \includegraphics[width=1\linewidth]{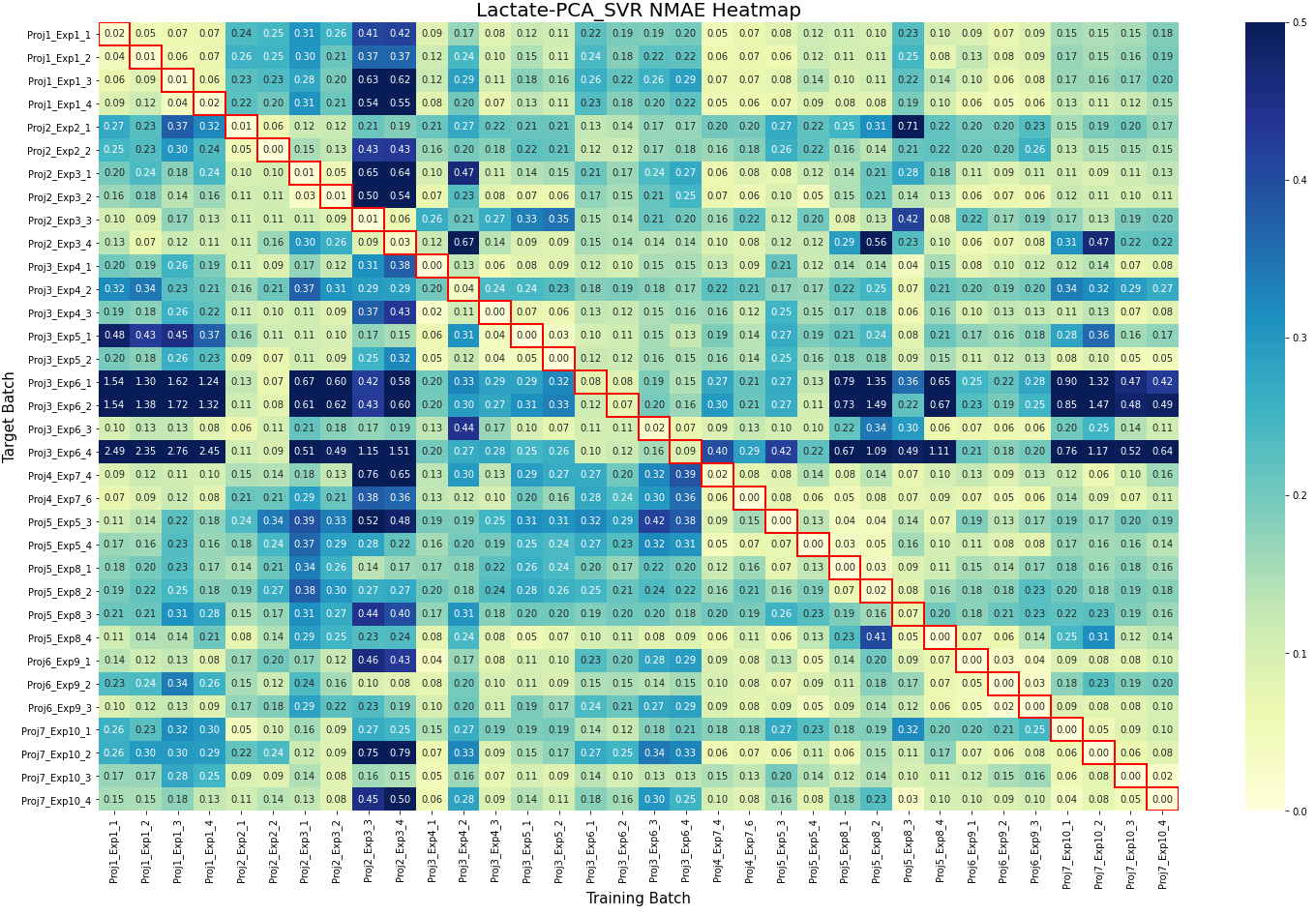}
    \caption{Heatmap of NMAE for predicting Lactate}
    \label{fig:Heatmap-Latacte}
\end{figure}

For the CSL Experimental dataset, although all runs are done in a 5-L scale bioreactor, the transferability of bioreactor runs is limited by differences in cell lines, feed media, and base media, as changes in them could significantly change the Raman signature of the bioreactor runs. These limitations lead to a different mapping relationship between Raman spectra and offline measures, limiting the models' transferability to another batch. Such limitations on transferability can be identified easily from our transferability analysis when the results are visualised as a heatmap. For example, figure~\ref{fig:Heatmap-Latacte} visualises the NMAE of PLSR models for the Lactate prediction task as a heatmap. The horizontal axis of the heatmap indicates the batch used for training the model, e.g. the training batches. The vertical axis indicates the batch used for evaluating the model, e.g. the target batches. The Exp\# in the batch name indicates the name of the experiments. Each experiment involves 1-4 bioreactors run in parallel with similar experimental settings, feed, and fill media. The colours indicate the performance (NMAE) of the model on the target batch with a colour scale ranging from 0 to 0.5.  The diagonal axis of the heatmap is highlighted with red lines. Models along the diagonal axis, which were trained and evaluated on the same dataset and are included in the graph for validation and reference, performed well as expected.
Furthermore, there are various pockets (closed by orange-coloured boxes) along the diagonal axis with clearly lighter colours than the models outside the pocket. This is because both the y and x-axis are sorted by experiment names, and projects that shared the identical Exp\# generally had similar experiment configuration, feed, and fill media, which means models usually are transferable amongst these batches, which leads to significantly better performance compared to models built using data from batches with different Exp\#, creating a ``pocket" of lighter colour cells within the heatmap. The existence of these pockets highlights that these bioreactor-run data are heterogeneous, limiting the transferability of the models built using them. Although good performance can still be achieved by the models within the same pockets, e.g. with a warm start, in reality, each bioreactor run will last for two weeks, so the runs within the same pocket are performed in parallel to save experimentation. For that reason, those results are actually not achievable in real time.

\begin{figure}[H]
    \centering
    \includegraphics[width=1\linewidth]{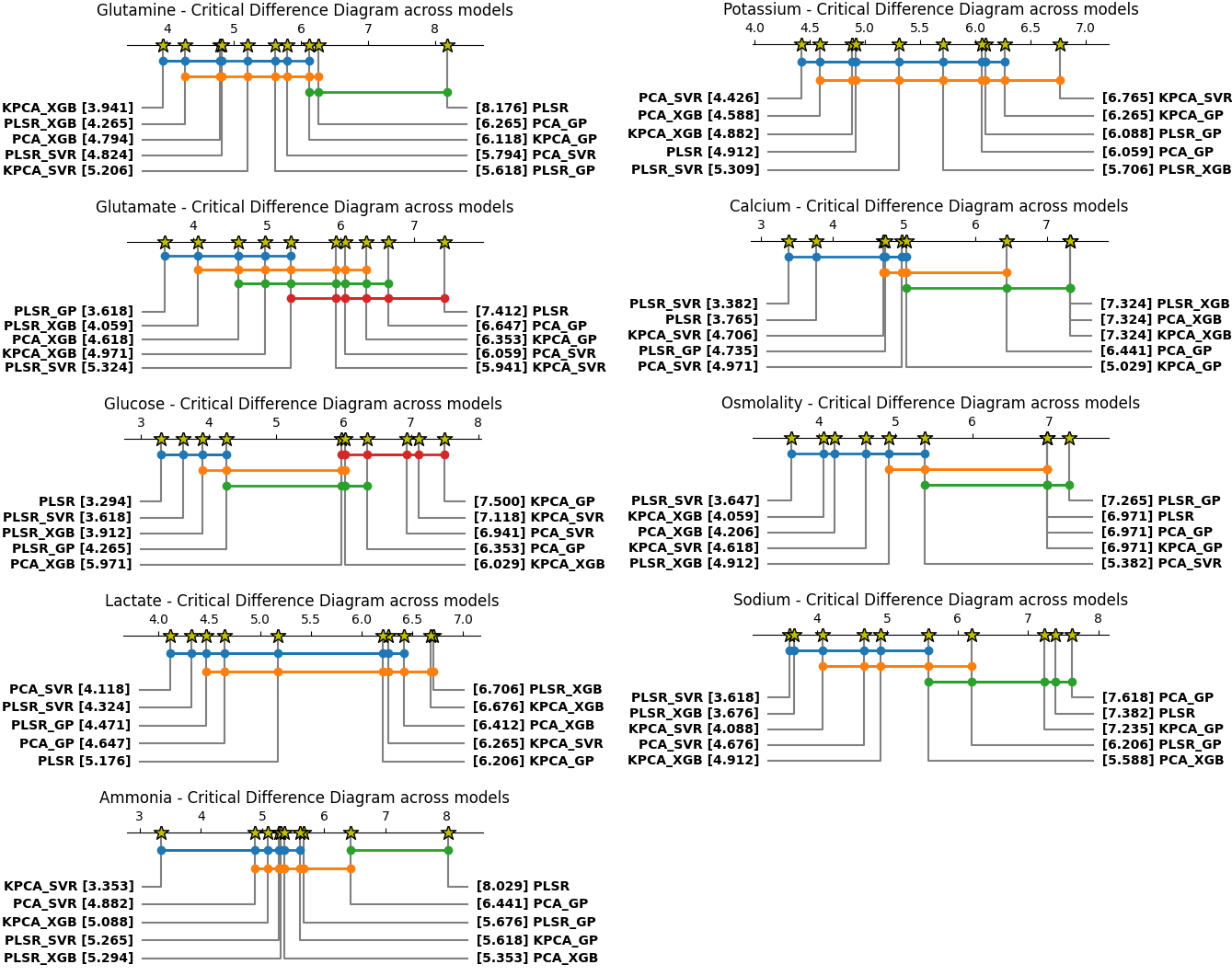}
    \caption{Critical Difference Plot for Offline Measurements}
    \label{fig:TransferabilityAnalysisCDP}
\end{figure}

Apart from the homogeneity between the training and target batch, \textbf{the complexity of DR and prediction model} could also be a limiting factor for model transferability. To demonstrate this, we visualised the test results with CDPs shown in Fig.~\ref{fig:TransferabilityAnalysisCDP}. Across all the CDPs, PLSR as the DR method has shown great results, and the best performing (highest average rank) model for 5 out of 9 offline measurements uses PLSR as the DR method. The other two DR methods, PCA and KPCA, are each used by the best-performing model for two offline measurements. Three insights could be drawn from this result. First, both PLSR and PCA are linear DR methods, whereas KPCA is nonlinear. The stronger performance of the simpler linear methods suggests that less complex DR techniques may be preferable when data is limited. Second, PLSR is the only supervised DR method among the three, indicating that supervised DR may be more effective for Raman-based soft sensor development. This is likely because each target variable has a unique Raman signature composed of specific wavenumbers. Unlike unsupervised DR, which compresses all information without considering relevance to the target, supervised DR selectively retains meaningful features, reducing noise and improving predictive performance. Finally, Glutamine and Ammonia were the only two offline measurements where KPCA outperformed other DR methods, suggesting that nonlinear relationships in Raman spectra may be particularly important for their prediction.

Across all prediction models, SVR was used as the best-performing model for 6 out of 9 offline measurements. The other three prediction models, GP, XGB, and PLSR, were each used by the best-performing model for one offline measurement only. The key distinction between SVR and the other model is that it is a Structural Risk Minimisation (SRM) \citep{vapnik1974teorija} model rather than an Empirical Risk Minimisation (ERM) \citep{Nips1991_ff4d5fbb} model. These results may suggest that SRM models are better for building Raman-based soft sensors when data is limited. This is unsurprising as SRM explicitly minimises training error and overfitting risk by controlling model capacity, making them less prone to overfitting. 

A common theme that can be highlighted across both DR and prediction models is that, under a limited data scenario, \textbf{low complexity models or models with complexity explicitly controlled, e.g. SRM models, perform better}. This result aligns with our finding from the literature review - models with more complexity would require more data points to be accurate. Thus, while data homogeneity plays an important role in the model's batch-to-batch transferability, the model complexity should also be carefully considered to maximise the ML models' batch-to-batch transferability.   

\subsubsection{IndPenSim datasets}
\begin{figure}[H]
    \centering
    \includegraphics[width=1\linewidth]{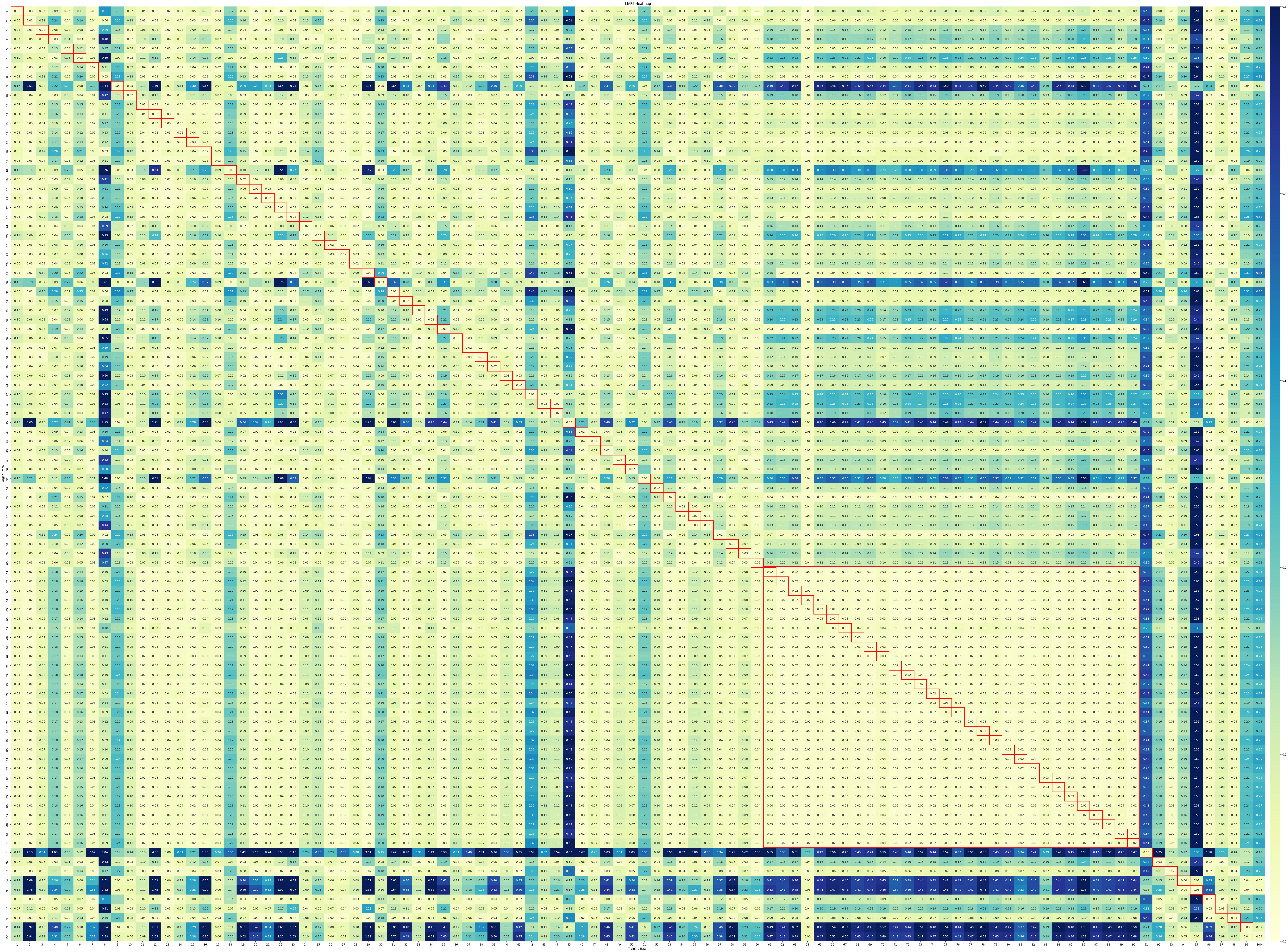}
    \caption{NMAE Heatmap for Peniciling Concentration Prediction - Training Batch on X-axis, Target Batch on Y-axis}
    \label{fig:HeatmapIndPenSim}
\end{figure}

\
The general transferability for the IndPenSim dataset is much better than the CSL dataset, as the heatmap (Fig.~\ref{fig:HeatmapIndPenSim}) has many more light-colour cells, meaning overall better batch-to-batch model transferability. This result is expected as the simulation basis for all batches is the same, and the only significant difference is their control strategies. Additionally, it is worth noting that the batches (batches 91 to 100, last ten rows in the heatmap) simulated with faults have much lower batch-to-batch transferability either as a training batch or a target batch. On the other hand, the large pocket highlighted by the red box in the bottom right corner of the heatmap performed significantly better than the areas around it. The models in this pocket were trained based on batches with Raman-based control and showed better transferability and consistency.

\begin{figure}[H]
    \centering
    \includegraphics[width=1\linewidth]{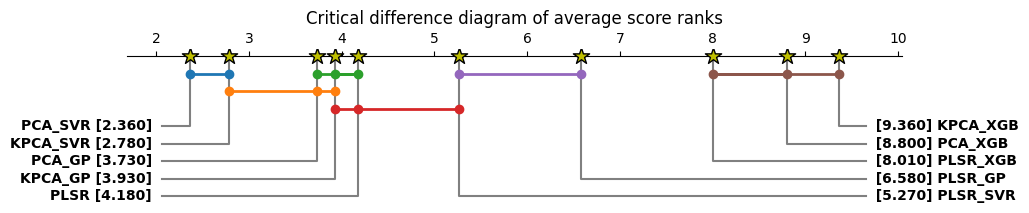}
    \caption{Critical Difference Plot for comparing the performance of ML models for Penicillin Concentration Prediction}
    \label{fig:IndPenSimCDP}
\end{figure}

Furthermore, the CDP shown in Fig~\ref{fig:IndPenSimCDP} indicates significant model differences. Overall, the two SVR models with PCA and KPCA performed the best and the second best, respectively. Across the models with the same DR methods, SVR seems to outperform GP consistently, while all XGB models ranked last, regardless of the dimensionality reduction methods. This is consistent with our findings from the CSL dataset, suggesting that SRM models like SVR perform better under limited data scenarios, while overly complex models such as XGB are unsuitable for limited data scenarios. Across all dimensionality reduction models, PCA appears to be the best choice, followed by KPCA and then PLSR, and the differences in performance between PCA and KPCA are not statistically significant. This result is interesting as PLSR is the best DR method for the CSL dataset. The poor performance of PLSR as a DR method on the IndPenSim dataset is likely due to the fact that PLSR models are unable to adapt to changes in later phases of the penicillin fermentation process, as shown in Fig.~\ref{fig:IndPenSim_PLSR_x_SVR}, which will be discussed in more detail in section~\ref{sec:Training Strategy}.

\subsubsection{AstraZeneca Dataset}
\begin{figure}[H]
    \centering
    \includegraphics[width=0.75\linewidth]{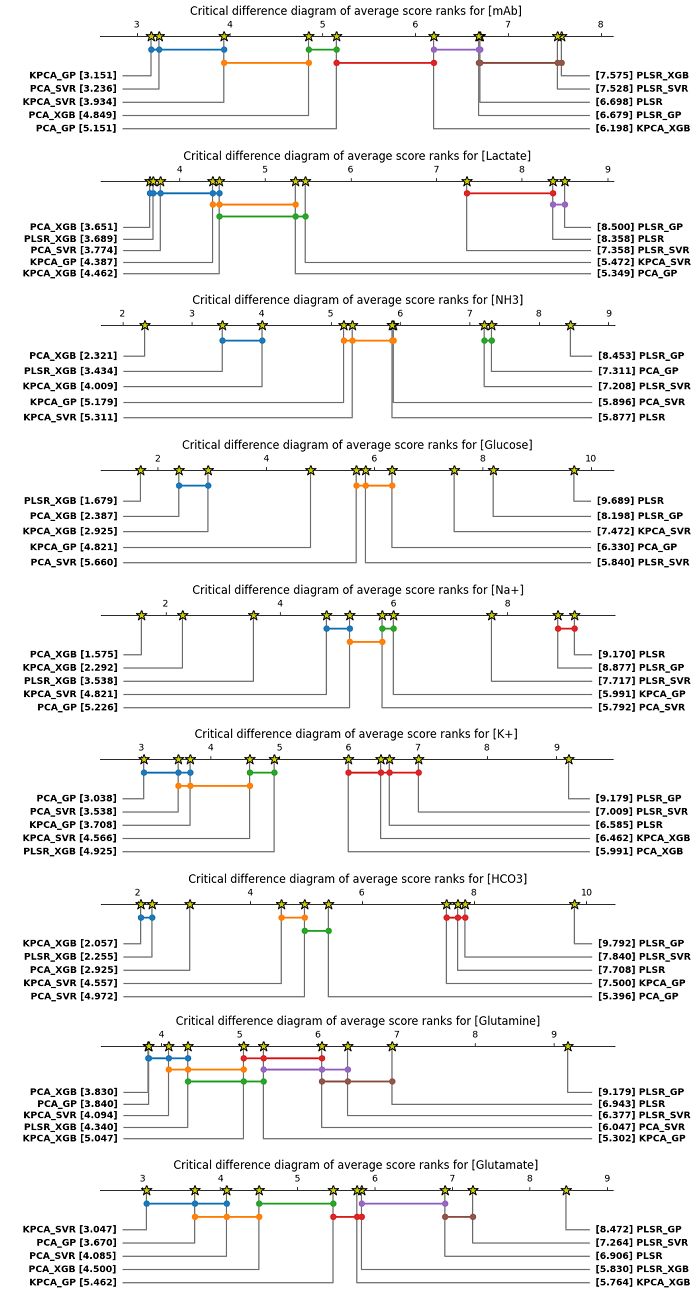}
    \caption{Critical Difference Plot for comparing the performance of ML models for Offline Measurement Prediction}
    \label{fig:AstraZeneca-Transferability-Analysis}
\end{figure}

\begin{figure}[H]
    \centering
    \includegraphics[width=0.7\linewidth]{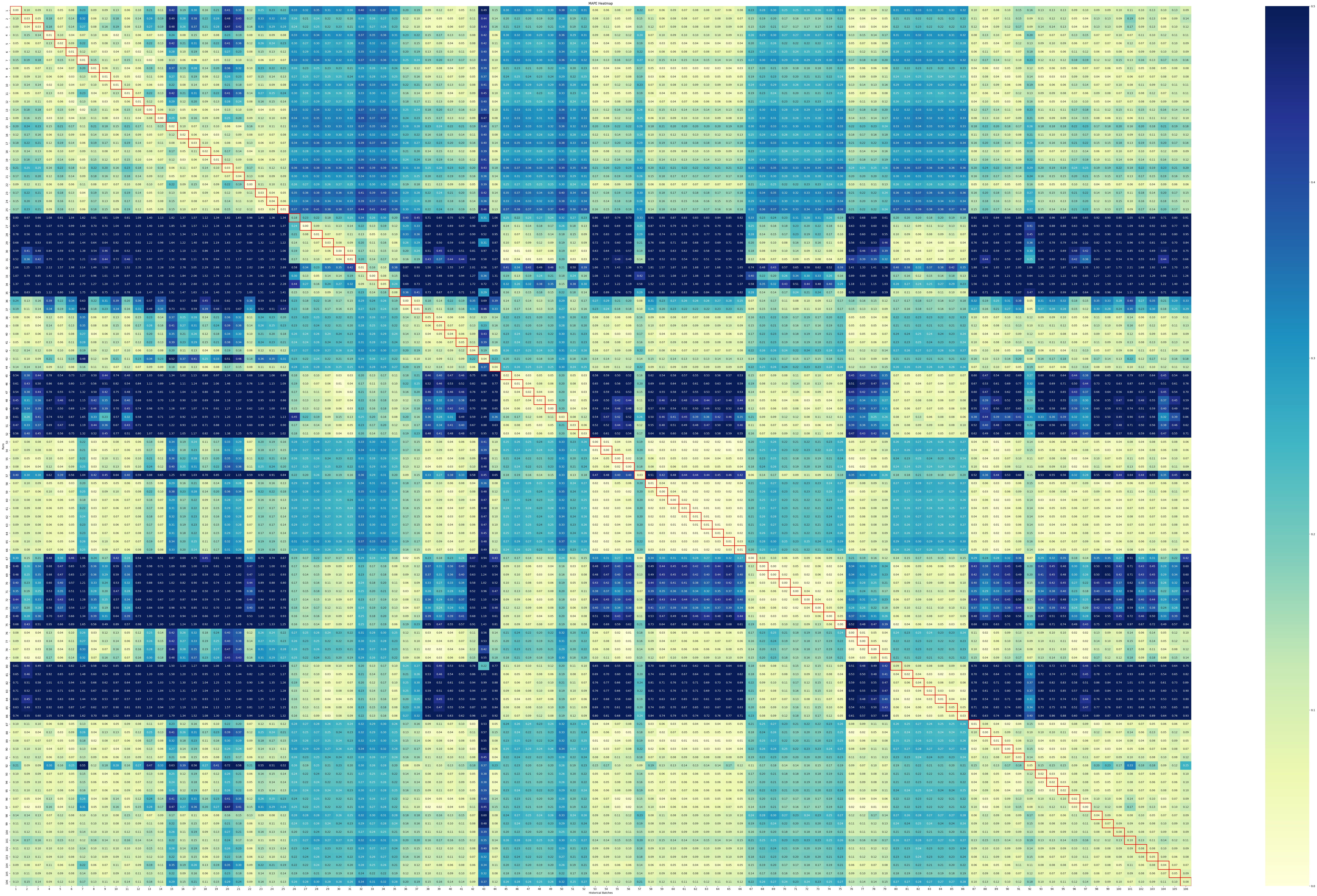}
    \caption{NMAE Heatmap for mAb Prediction - Training Batch on X-axis, Target Batch on Y-axis}
    \label{fig:Heatmap-AZ}
\end{figure}

The transferability analysis for AstraZeneca data is carried out with a similar approach, with results visualised as CDPs (Fig~\ref{fig:AstraZeneca-Transferability-Analysis}) and a heatmap (Fig~\ref{fig:Heatmap-AZ}). The key distinction for the benchmarking on AstraZeneca data is that we perform one-step-ahead prediction instead of monitoring. Thus, feedback is provided as lagged observations to pre-trained models, which is not the case for the other two datasets. 

Overall, XGB models have demonstrated great performance and achieved the highest average rank for 6 out of 9 offline measures, which is surprising as they did not perform well on the other two datasets. This might be because input features are now made of nine different lagged offline measurements instead of Raman spectra, which is highly sparse, as many lagged offline measurements will be imputed as zero at the early stage of the bioreactor run. For high-sparsity data, XGB, as a tree-based model, naturally handles sparse input features, while the other ML models may struggle. For example, PLSR consistently ranked at the bottom as it faces convergence issues when learning from highly sparse data. Furthermore, each lagged offline measurement could have a different relationship with the target variable, leading to a heterogeneous feature space. This makes the XGB a great model for this task because the boosting mechanism of XGB allows it to flexibly learn from features with different distributions by constructing new trees. In this case, the relationships between the input features and target variables are more complex than those in the other two datasets. Hence, more complicated models like XGB have shown better performance. Thus, the results are still reasonable and consistent with our previous conclusion that the model complexity should match the complexity of the data and tasks.
The heatmap for mAb prediction models is shown in Fig~\ref{fig:Heatmap-AZ}.  Similar to the two datasets, it has also demonstrated multiple "pockets", highlighting groups of bioreactor runs that are more likely to be similar in experimental settings. Even though AstraZeneca did not release any descriptions or meta-information about the individual cell lines and experiment settings used for each run, this result is interesting because the pockets from the other two datasets likely resulted from variations in experiment settings, such as cell line and feed media, which can impact the Raman spectrum. However, in this case, our input features only consist of lagged offline measurements, which indicates that the behaviour of the target offline measurements and their relationship with lagged offline measurements are also likely impacted by the experimental setting. An example is that the timing of feeding may influence the behaviour of the glucose value, which in turn affects the values of other offline measurements, such as lactate, as it is a metabolite of glucose. While this result is within expectation, it nonetheless provides valuable insights into the problem.       
\subsection{Meta Analysis}
The transferability analysis highlighted the challenges of cold-start scenarios, and it is in our interest to understand the meta-features that drove the differences in homogeneity between the data. In this section, we will perform a meta-analysis on the benchmark datasets to explore the potential meta-features that could impact the homogeneity between datasets. The findings from the meta-analysis will enable us to curate training datasets that are homogeneous with the new bioreactor run or include these meta-features as part of the training data, which may improve the accuracy of the soft sensor. Note that no meta-features have been provided for AstraZeneca data. Thus, a meta-analysis for the AstraZeneca dataset was not performed.    

\subsubsection{CSL datasets}\
The two most important meta-features available in the CSL dataset are feed media and base media, which could significantly impact the Raman signature of the offline measures. Thus, we have performed a meta-analysis based on the feed and fill media used for the bioreactor runs. Firstly, we divided the individual models trained during transferability studies into one of the following groups:
\begin{enumerate}
    \item Model trained on data that shared the \textbf{same feed media} as the target batch
    \item Model trained on data that shared the \textbf{same fill media} as the target batch
    \item Model trained on data that shared the \textbf{same fill and feed media} as the target batch
    \item Model trained on data that \textbf{do not share the fill or feed media} as the target batch  
    \item Model trained on data from bioreactor runs that ran in parallel with the target batch using a \textbf{similar experiment setting} (the same Exp\#). 
\end{enumerate}

\begin{figure}[H]
    \centering
    \includegraphics[width=1\linewidth]{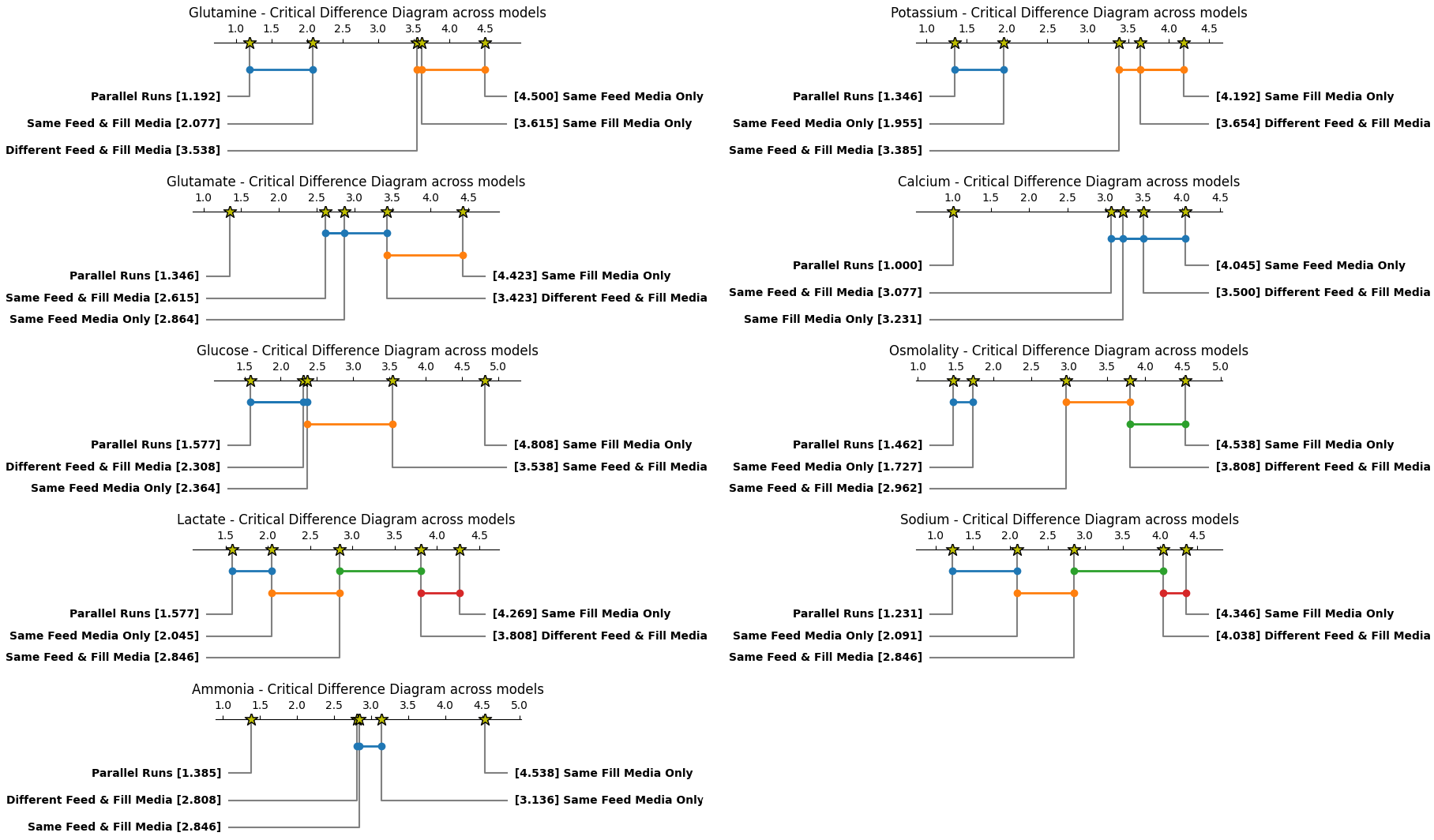}
    \caption{Critical Difference Plots for different model groups}
    \label{fig:CDPMetaAnalysis}
\end{figure}

Then, we calculated the average ranks of all benchmark models on all runs following the leave-one-batch-out testing protocol as discussed in \ref{sec:ExpDesign}. Statistical testing was performed using Friedman and Nemenyi's post hoc tests to test whether there were significant differences in the model between these groups.  The results are visualised as critical difference plots (CDP) with average rank in squared brackets shown in (Fig~\ref{fig:CDPMetaAnalysis}). The results are as expected: the model trained on data generated from parallel runs with similar experiment settings achieved the highest average rank for all offline measurements and consistently significantly outperformed models trained on data from runs with different feed and fill media. Thus, having no access to runs with similar experiment settings, e.g. a "cold start" scenario, adds significantly more challenges to building accurate Raman-based soft sensors. It is also worth noting that if we put the parallel run models aside, the models with the same feed media seem to perform the best for most offline measurements, except for Ammonia, Glutamine, and Glutamate. For them, models trained with the same feed \& fill media performed the best.
On the other hand, models trained on data with different feed \& fill media or with the same fill media only generally performed the worst across all offline measurements, suggesting feed media has a more significant impact on the homogeneity of the data than fill media. While the results are reasonable, it is essential to note that this analysis is just a preliminary one-way analysis of the fill and feed media, which does not account for differences such as feeding plans and set points. It is possible that other uncontrolled differences between runs were causing the inferior performance of the model, despite using the same feed and fill media for Lactate. More analysis would be required to identify the drivers for these results, which is outside the scope of this benchmark report. Nonetheless, this meta-analysis on CSL data provided preliminary insights into how meta-features such as fill and feed media may impact the homogeneity of the data, which can significantly affect the batch-to-batch model transferability. Future studies on the development of soft sensors could consider building a meta-learning model that leverages these meta-features to initialise the hyperparameters and parameters of the soft-sensor models, thereby handling the cold start scenario more effectively.    

\subsubsection{IndPenSim datasets}
\begin{figure}[H]
    \centering
    \includegraphics[width=0.8\linewidth]{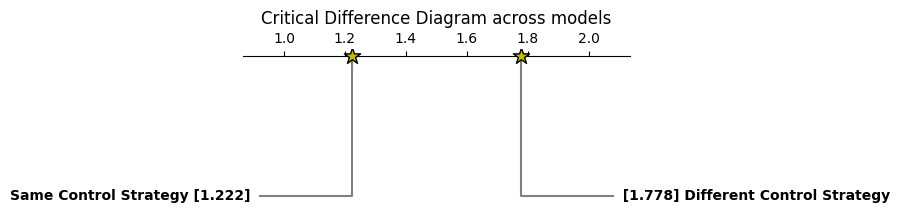}
    \caption{Critical Difference Plot for models trained using data from different control strategies}
    \label{fig:IndPenSimMetaAnalysis}
\end{figure}

\
Similar to the CSL dataset, we are also interested in identifying meta-features that could potentially impact the transferability of the data and models. However, unlike the CSL data, the IndPensim data was designed to investigate the impact on penicillin production from different control strategies. Thus, the primary difference between the various runs lies in the control strategy. Therefore, we have divided the individual models trained during transferability studies into one of the following groups:
\begin{enumerate}
    \item Model trained on data shared the \textbf{same control strategy} as the target batch
    \item Model trained on data with \textbf{control strategy different to} the target batch  
\end{enumerate}
Then, we performed the same statistical test as before to generate the CDP (Fig.~\ref {fig:IndPenSimMetaAnalysis}) and determine whether these model groups can identify significant differences. The test results are as expected: the model trained on data generated with the same control strategy significantly outperformed the model trained on data with a different control strategy. This result suggests that not only will feed and fill media impact the data homogeneity but it could also be influenced by a control strategy, indicating that a more comprehensive meta-analysis would be required to capture all factors that would impact the data homogeneity and batch-to-batch model transferability.   

\subsection{Training Strategy}
\label{sec:Training Strategy}
As highlighted from the transferability analysis, batch-to-batch model transferability could be significantly impacted if there is a lack of relevant data in the training set. This issue can be mitigated by allowing the ML model to learn from historical offline measurements acquired from the target batch, such as feedback. This way, the ML model can adapt to the target data distribution as the bioreactor run progresses, thereby improving model performance over time. However, given that the offline measures we are trying to monitor are generally expensive or challenging to measure, the feedback is expected to be provided at a limited volume and low frequency. Thus, exploring ways to effectively leverage this limited feedback, along with the limited historical data, is crucial.

The most straightforward way to adapt models to learn from new feedback is to combine it with historical data and then retrain the ML model using the combined data. However, this retraining strategy is not ideal as all data points carry the same weights. Still, the new feedback from the target run should carry more weight than the historical data from different bioreactor runs, resulting in a slow model adaptation to the new feedback. Thus, as discussed in the methodology section, we will also examine the effectiveness of two alternative retraining strategies - \textbf{JITL} and \textbf{OL}, which may improve the speed of model adaptation and performance. 

\subsubsection{CSL Dataset}
\begin{figure}[H]
    \centering
    \includegraphics[width=0.725\linewidth]{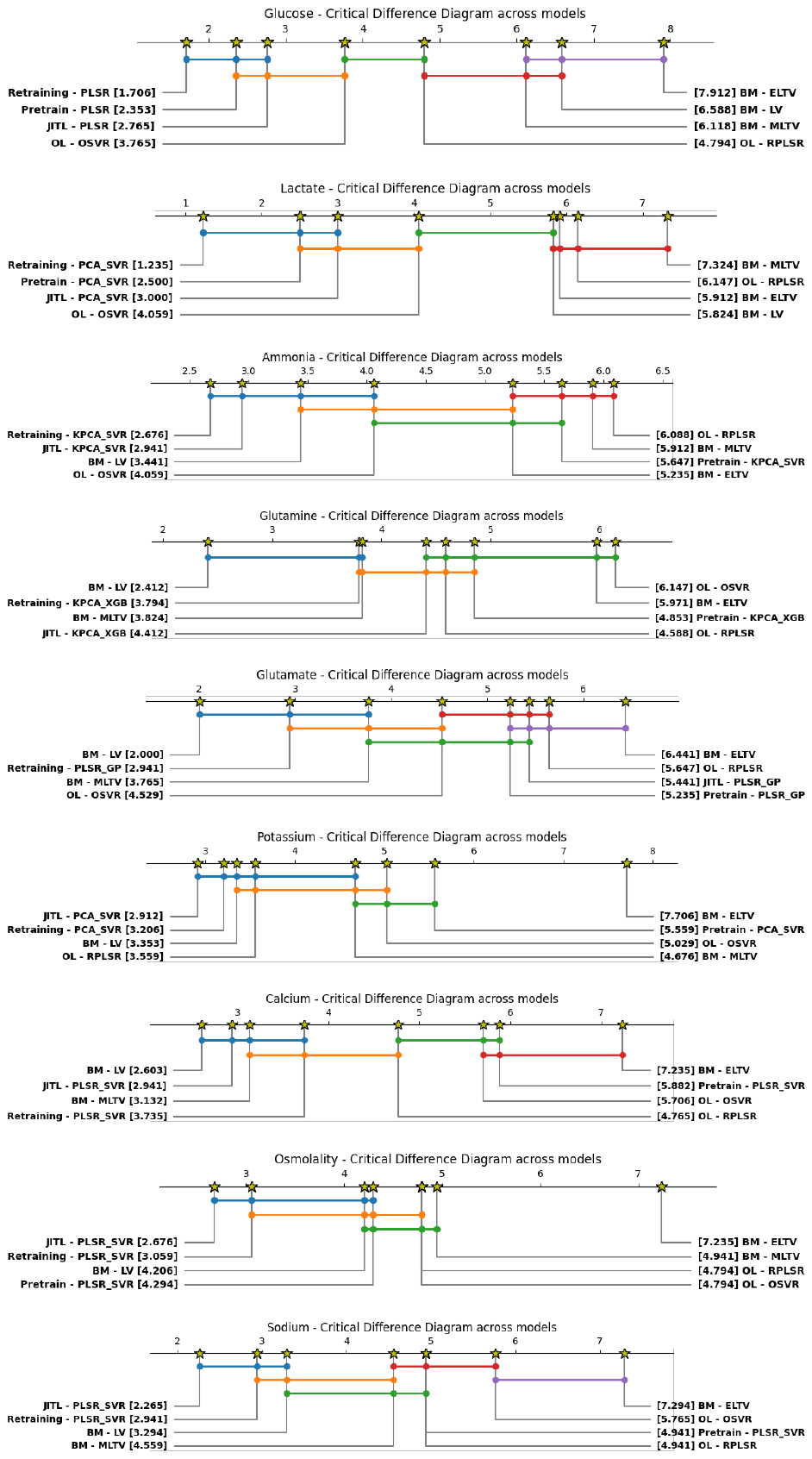}
    \caption{Critical Difference Plots for Comparing Training Strategies for CSL Dataset}
    \label{fig:CSL-OL}
\end{figure}

\
Fig.~\ref{fig:CSL-OL}  shows the CDPs for nine different offline measurements and compares the average ranks between training strategies, including No Training (Baseline Models), Retraining, Pretraining, OL, and JITL. The first important finding is that retraining consistently outperforms pre-training and has shown significant improvement in most offline measures. This result suggests that incorporating just a few additional data points from the new runs could significantly enhance model performance, underscoring the importance of learning from recent feedback when the historical dataset is limited. 

Secondly, both retraining and JITL are more effective than pretraining and OL, with each achieving the highest average rank for three offline measurements. This result is somewhat surprising, as the results shown in \cite{Tulsyan2019} suggest that JITL should dominate the global retraining model. This result may be explained by the fact that, although our historical data is limited, it is more homogeneous when compared to the data used in \cite{Tulsyan2019}, which had more than 7,000 data points generated from runs with different cell lines and experimental settings. Thus, the performance of the global models in that paper was relatively poor as a baseline model due to data heterogeneity, which allowed JITL models to demonstrate significant improvement over them. The same level of improvement on the global model is not possible with our relatively more homogenous data. More supporting evidence can be found in Section~\ref {sec:validation run}, where JITL consistently outperforms the global retraining model when evaluated using two validation runs with heterogeneous data distribution, e.g., under a cold-start scenario.

The final important finding is that, somewhat unexpectedly, the baseline model, which uses the last observed value as the prediction, achieved the highest average rank for three offline measurements. This is likely because offline measurements, such as Sodium and Calcium, typically change only once or twice throughout the run, making the most recent value a surprisingly reliable indicator. This finding highlights the importance of integrating historical offline data with Raman signals. Future work should aim to develop frameworks that can effectively learn from heterogeneous inputs—such as time-stable offline measurements and dynamic Raman data—which may significantly enhance predictive performance compared to models based solely on Raman spectroscopy.

Note that we did not optimise the hyperparameter $k$ for JITL and set it to 30 to replicate the implementation in \cite{Tulsyan2019}, which might not be the best choice of $k$ for our dataset. Similarly, OL learnt all historical data in a fixed order in this benchmark study. However, OL might benefit from learning from fewer historical datasets and in different orders, allowing it to adapt more quickly and robustly to changes in the new runs. Therefore, JITL and OL may perform better if more optimisation is done.   
\begin{table}[H]
    \centering
    \includegraphics[width=1\linewidth]{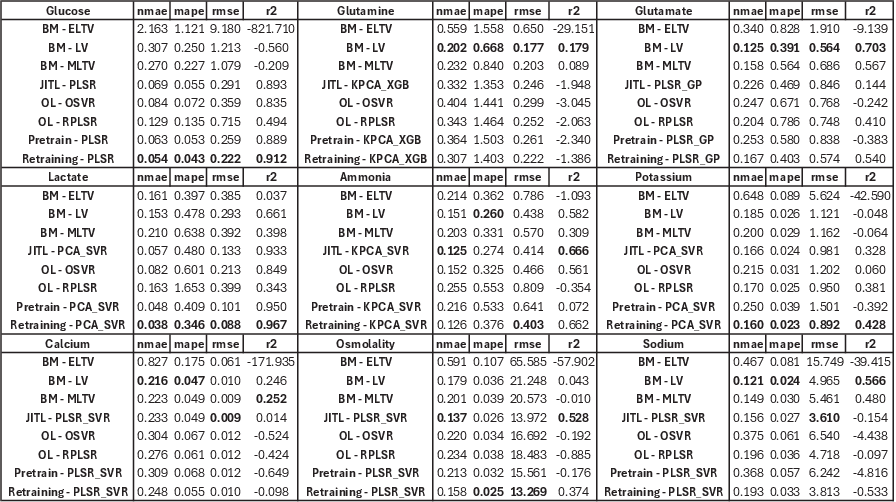}
    \caption{Performance Metrics for Comparing Training Strategies}
    \label{fig:Metrics-CSL-OL}
\end{table}

Lastly, as shown in Tabble~\ref{fig:Metrics-CSL-OL}, the Raman models have improved their performance significantly by having access to online feedback and retraining. Noticeably, Ammonia, Osmolality, and Potassium all performed better than the baseline models after retraining. However, as shown in the CDPs, Raman models still struggle to significantly outperform the baseline methods for offline measures other than Lactate and Glucose. These results highlight the importance of incorporating lagged observations as part of the input features in ML models for the monitoring task, as using only Raman data is insufficient to monitor many offline measurements accurately. It is crucial to develop new methods for training soft sensors that utilise multiple sources of information, including lagged observations, as well as Raman data, as their input features. 
\subsubsection{IndPenSim Dataset}
\begin{figure}[H]
    \centering
    \includegraphics[width=1\linewidth]{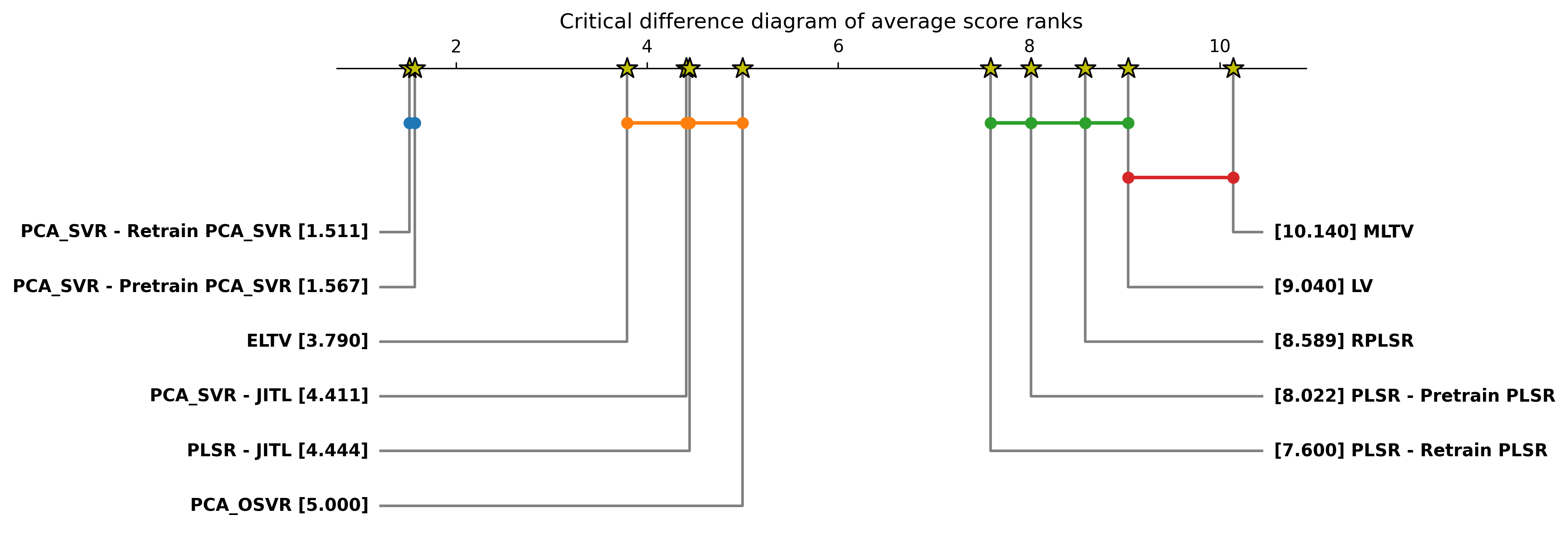}
    \caption{Critical Difference Plots for Comparing Training Strategies on IndPenSim Dataset}
    \label{fig:IndPenSim-OL}
\end{figure}

For the IndPenSim dataset, the best training strategy remained batch retraining, as shown in CDP plots in Fig.~\ref{fig:IndPenSim-OL} and performance metrics in Table.~\ref{fig:Metrics-IndPenSim-OL}. Both Batch retraining and pretraining performed significantly better than the OL and JITL models. Furthermore, unlike the results from the CSL dataset, there are no significant differences between retraining and pre-training models. These results are expected, as IndPenSim data is significantly more homogenous than the CSL dataset (as shown from the heatmap), which reduces the importance of the new feedback, leading to a lower ranking of online learning and JITL models, as well as minimal differences between retaining and retraining models. 

\begin{table}[H]
    \centering
    \includegraphics[width=0.5\linewidth]{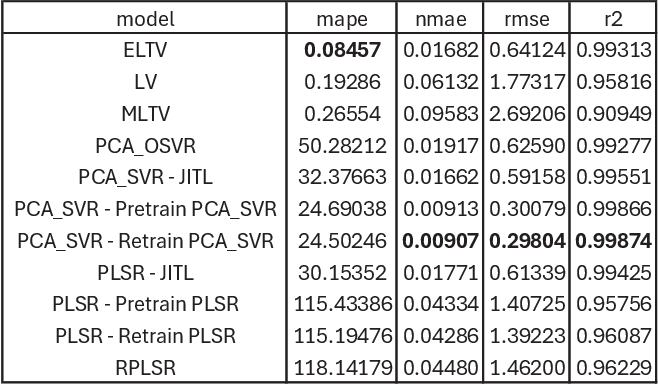}
    \caption{Performance Metrics for Comparing Training Strategies for IndPenSim Data}
    \label{fig:Metrics-IndPenSim-OL}
\end{table}

\begin{figure}[H]
    \centering
    \includegraphics[width=1\linewidth]{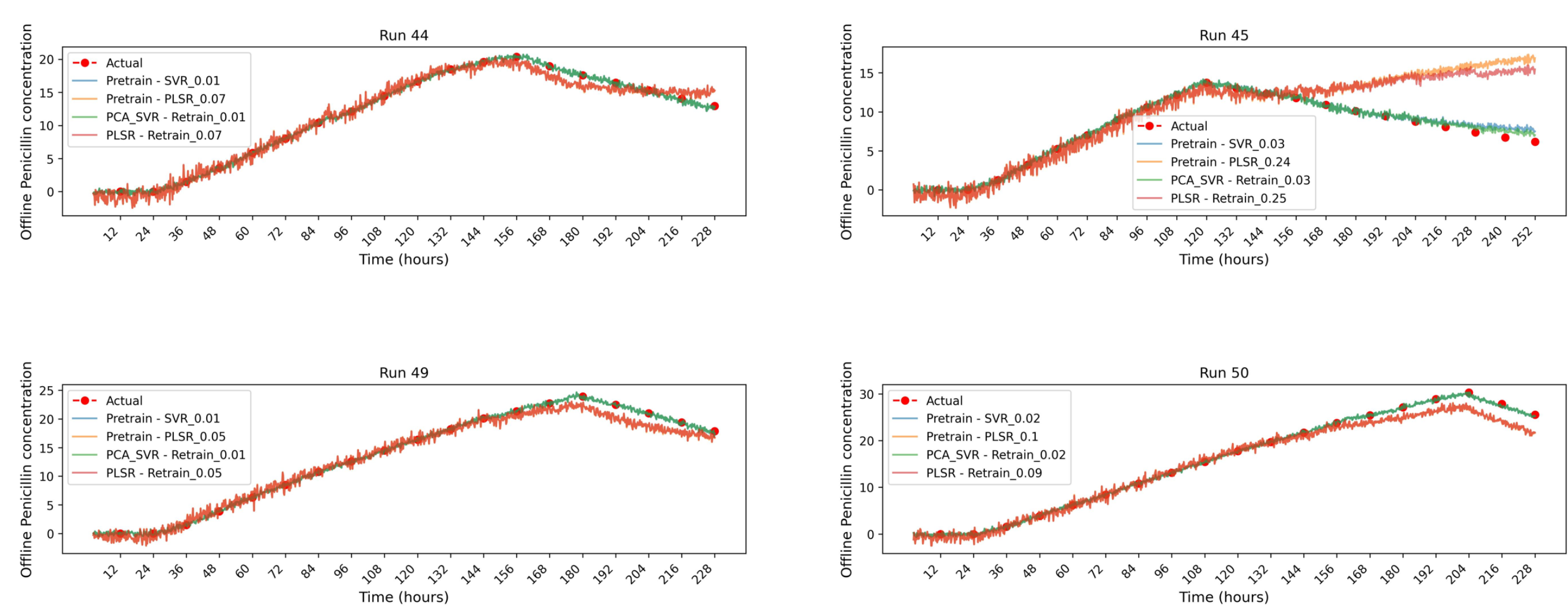}
    \caption{Comparing Model Predictions between PLSR and SVR models}
    \label{fig:IndPenSim_PLSR_x_SVR}
\end{figure}

Across the two OL models, OSVR outperformed RPLSR significantly. This is because PLSR-based models seem unable to pick up the changes in phases of the bioprocess and consistently deviate in the second half of the bioprocess, as shown in Fig.~\ref{fig:IndPenSim_PLSR_x_SVR}, which may suggest that the underlying mapping relationship might be non-linear as SVR with non-linear kernels were able to capture these changes. 

\begin{figure}[H]
    \centering
    \includegraphics[width=1\linewidth]{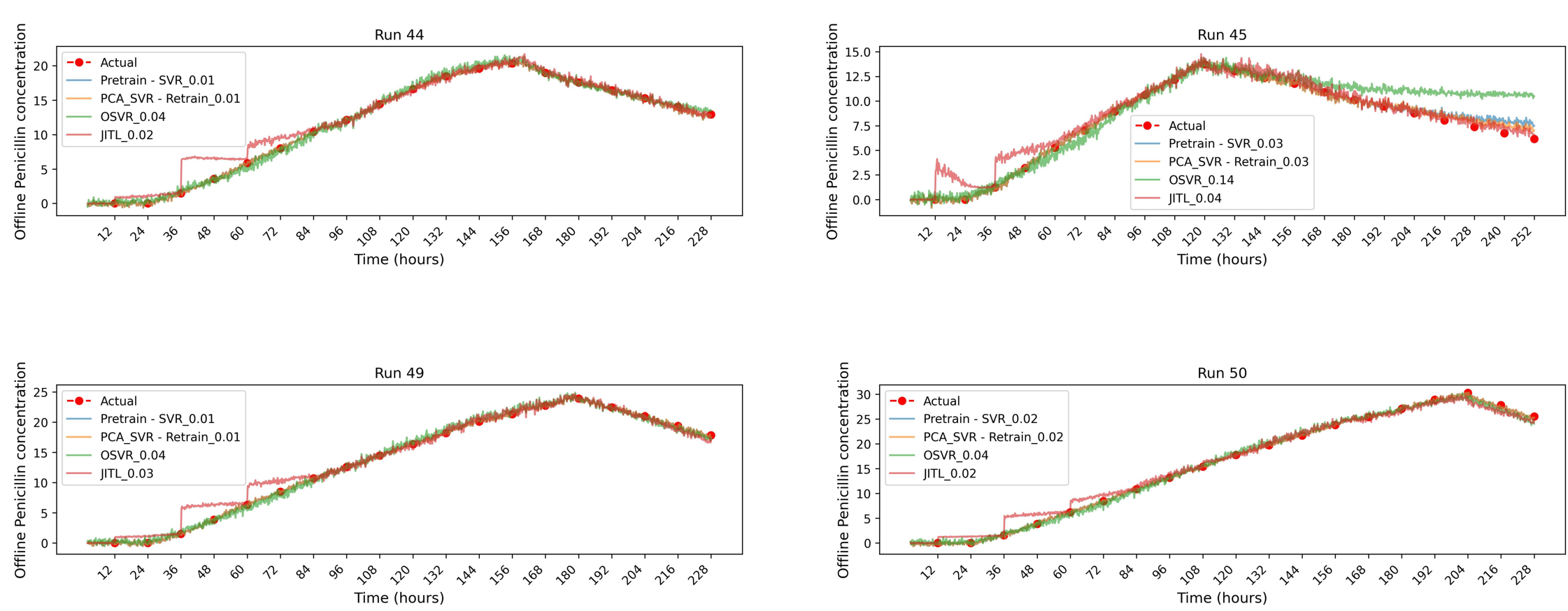}
    \caption{Comparing Training Strategies for SVR Models}
    \label{fig:IndPenSim_SVR}
\end{figure}

\begin{figure}[H]
    \centering
    \includegraphics[width=1\linewidth]{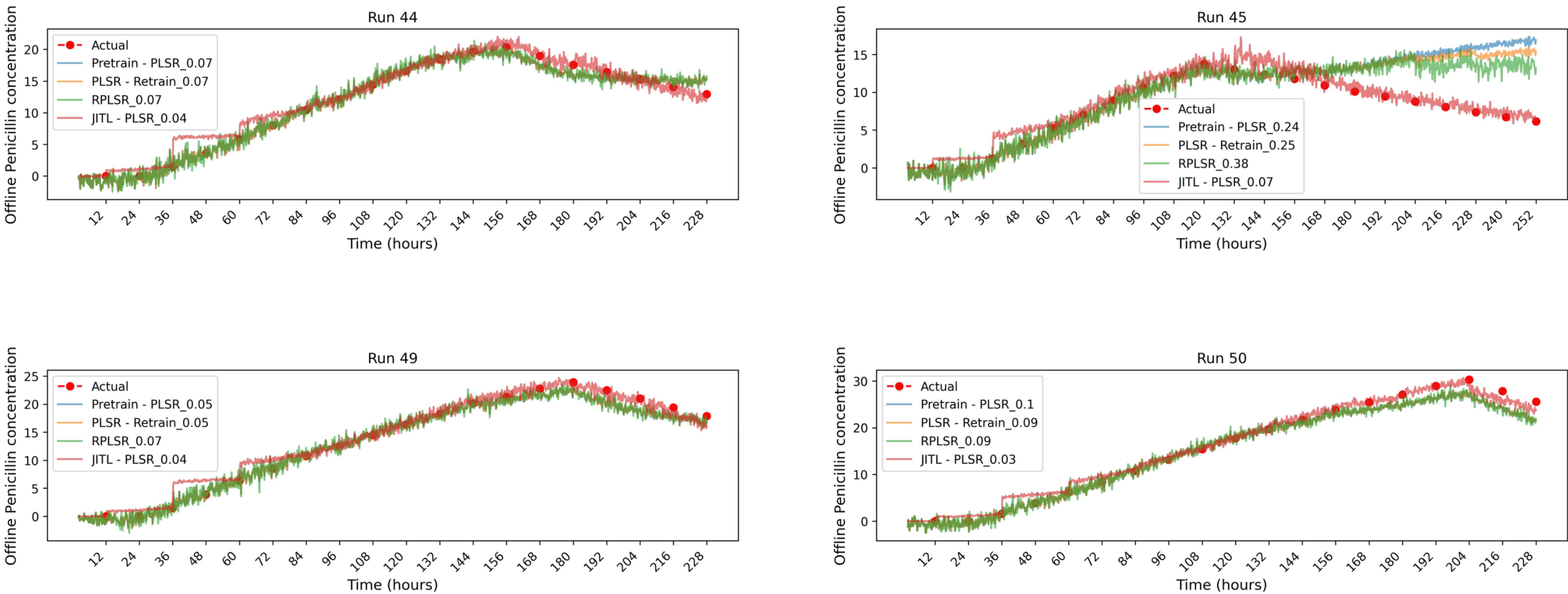}
    \caption{Comparing Training Strategies for PLSR Models}
    \label{fig:IndPenSim_PLSR}
\end{figure}

It is also worth noting that JITL shows better adaptation ability than other training strategies when data from the new run deviates from the historical runs, e.g. cold-start scenario. For example, as shown in Fig.~\ref{fig:IndPenSim_SVR} and Fig.~\ref{fig:IndPenSim_PLSR}, Run 45 has a distinctly different pattern compared to other runs, which caused both batch learning and OL methods to struggle to adapt to the changes quickly enough and perform poorly in the second half of the run for PLSR-based models and slightly worse for SVR-based models. On the other hand, JITL adjusted its prediction to follow the actual values quickly and produced significantly better predictions during the second half of the run. However, JITL produced notably worse predictions (flatlines) at an early stage of the bioprocess runs, leading to an inferior overall ranking compared to the batch learning models. This may be because the Penicillin Concentration values are close to 0 at the beginning of the run, resulting in a noisy Raman spectrum and causing the similarity measures to fail to retrieve the most relevant data points for training the JITL models. This issue may be addressed by using a similarity measure that is more robust to noise, which can be explored in future studies. 
\subsubsection{AstraZeneca Dataset}
\begin{figure}
    \centering
    \includegraphics[width=0.7\linewidth]{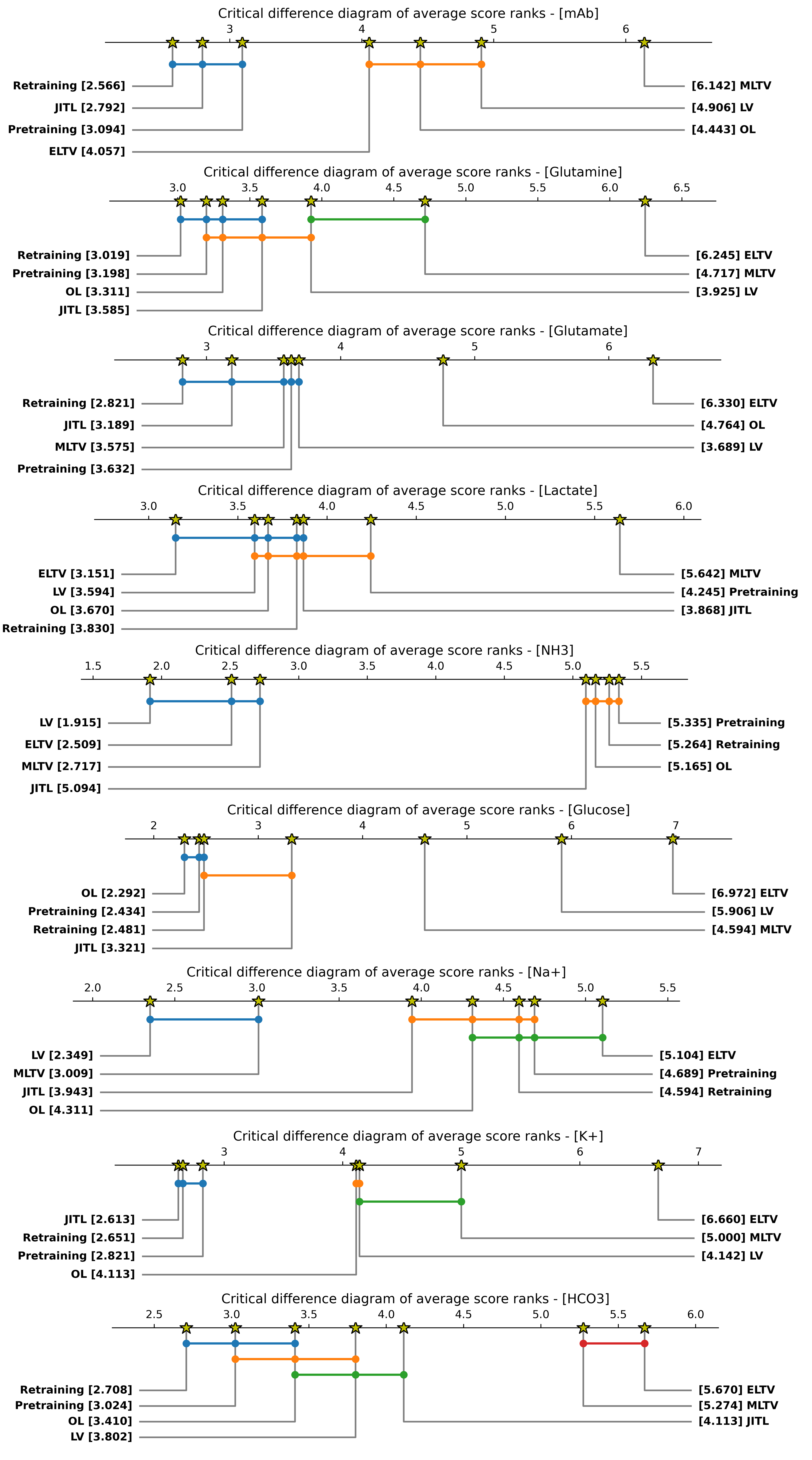}
    \caption{Critical Difference Plots for Comparing Training Strategies on AstraZeneca Dataset}
    \label{fig:OL_AZ}
\end{figure}

\
For the AstraZeneca dataset, the benchmark results on different training strategies are shown in Fig~\ref{fig:OL_AZ} and Table~\ref{fig:Metrics-AZ-OL}. In the transferability analysis, the XGB model performed best for the AstraZeneca Dataset. Thus, we used XGB with incremental learning as the OL model instead of PLSR or OSVR to better compare retraining strategies. Furthermore, MAPE is removed from the table because some offline values are normalised, and the values are close to or at 0, which causes erroneous MAPE. 

Overall, the results show that models trained on historically lagged offline measurements can produce reasonable next-step predictions and outperform the baseline methods for 6 out of 9 offline measurements, except for Lactate, Ammonia, and Sodium. Interestingly, our results from the CSL dataset show that Lactate, Ammonia, and Sodium can be predicted reasonably well based on Raman data, outperforming the baseline models. Thus, lagged offline measurements could potentially complement Raman data. 

For the six offline measurements in which ML models outperformed the baseline models, retraining is the best retraining strategy for 4 of them, while OL and JITL achieved the highest average rank for one of them each. Even in the case where OL and JITL performed the best, there is no significant difference between Retraining and JITL or OL. However, significant differences can be identified between OL and JITL. For mAb, Glutamate, and K+, JITL outperformed OL significantly, while OL outperformed JITL significantly for Glucose and HCO3. Thus, retraining appears to be the most robust training strategy across all offline measurements. At the same time, JITL and OL may achieve better, albeit non-significant, performance for specific offline measurements compared to retraining.

As discussed before, both JITL and OL could be potentially improved with some minor optimisation. For example, in this case, all features have the same weights when calculating similarity metrics for JITL. However, the time feature, ``Days Since Run Started", might be more beneficial for the model to learn from. Thus, if we could optimise the weights assigned to the time feature during similarity calculation, JITL may perform better than other training strategies. Therefore, these results on the AstraZeneca dataset are inconclusive, and further studies on both JITL and OL should be conducted to explore the potential of JITL and OL in more detail. 

\begin{table}[H]
    \centering
    \includegraphics[width=1\linewidth]{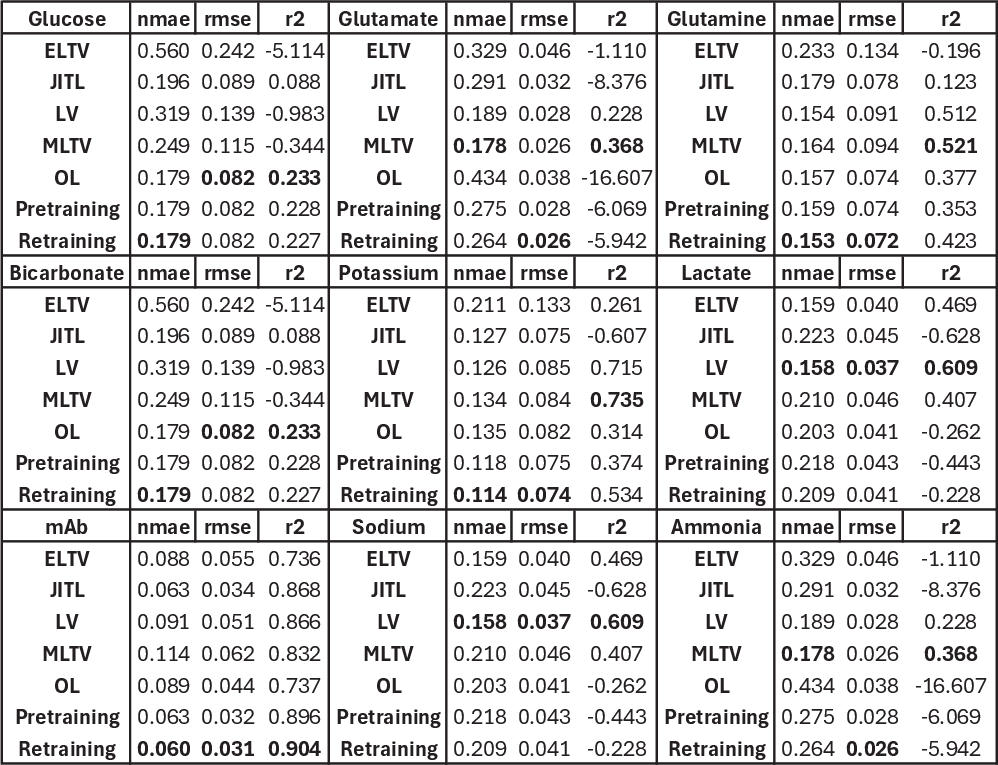}
    \caption{Performance Metrics for Comparing Training Strategies}
    \label{fig:Metrics-AZ-OL}
\end{table}

\subsection{Results on Validation Experimental CSL Data}
\label{sec:validation run}
Apart from the 34 runs of historical CSL data, we have also included two additional runs (A3 and A4), which were not used during the model optimisation and experimental process. These two runs are distinct from other historical runs in three significant ways. Firstly, the feed media used for them are entirely new and have not been used in any historical run. Secondly, these two runs used spread feeding instead of bolus feeding, a feeding strategy not used in any historical runs. Lastly, the cell line used in these two runs is also not used in historical runs. 

Performing real-time monitoring on these two runs can be seen as a cold-start problem, and it would be much ``colder" than the benchmark experiments we have done so far in the previous sections, given that these two runs have many unique characteristics unseen in the historical runs. Thus, they serve as a great validation set to test each training strategy's capabilities in handling the cold-start scenario. We have performed the test on Glucose, Lactate, and Ammonia. Overall, we have tested the following training strategies and models on these two runs (A3 \& A4):
\begin{itemize} 
    \item Online Learning - Recursive PLSR model for Glucose, Lactate, Ammonia
    \item Online Learning - Online SVR model for Glucose, Lactate, Ammonia
    \item JITL - PLSR for Glucose, PCA \& SVR for Lacate, and KPCA \& SVR for Ammonia
    \item Batch Training - PLSR for Glucose, PCA \& SVR for Lacate, and KPCA \& SVR for Ammonia  
\end{itemize}

Furthermore, to further test the adaptability of each model, we have divided the test into three scenarios: No Update, Daily Update, and Real-time Update (where feedback becomes available as soon as the offline measurements are taken). In the case of Daily Update, the offline measurements recorded by scientists will only become available at a fixed time of the day, simulating the potential data delay that could happen in practice. On the other hand, in the case of Real-time updates, feedback becomes available as soon as the offline measurements are entered into the system. In terms of training data, online learning models are trained on only one historical run with tuned hyperparameters, as training OL models on a large number of datasets might reduce their speed of adaptation on new runs, which is crucial for cold-start scenarios. In contrast, the batch training model was trained on all 34 historical runs, which JITL models also will have access to during training, similar to the global and JITL model setup in \cite{Tulsyan2019}.

The results of the performance of the models on the two validation runs are shown in Table~\ref{fig:ValRunMetrics}. As a result of the new fill/feed media and a new feeding strategy with daily interventions, as well as maintaining glucose concentrations at levels significantly higher than those in historical runs, the cell growth behaviours differed significantly from historical data. Consequently, batch retraining models did not capture the new cell growth behaviours and performed poorly. In contrast, online learning using pre-trained models on small datasets and JITL models—which build entirely new models using the 30 historical data points most similar to the latest samples—achieved better performance. JITL builds a new model for each new sample, allowing it to adapt more quickly to new cell growth behaviours than online ML models. Therefore, the performance of JITL models may be slightly better than that of online ML models. Lastly, as expected, models without any update performed poorly, highlighting the importance of having access to and adapting to the data from the new run.
\begin{table}[H]
    \centering
    \includegraphics[width=1\linewidth]{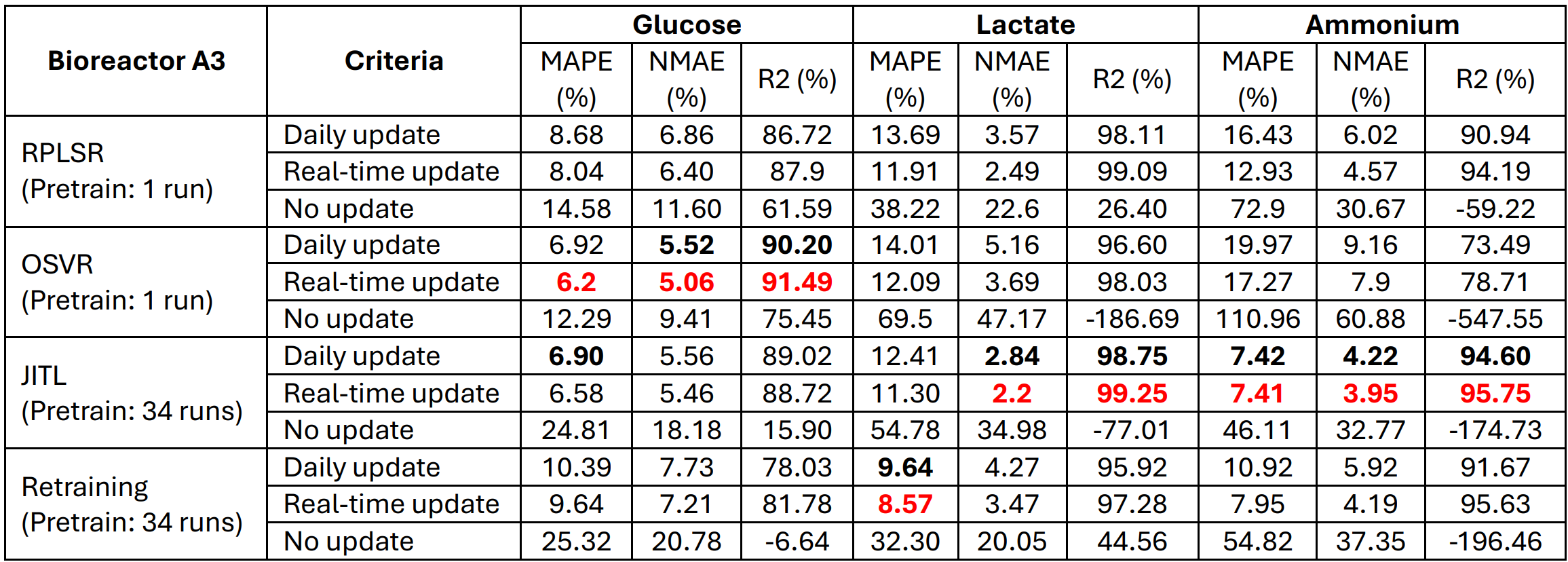}
    \includegraphics[width=1\linewidth]{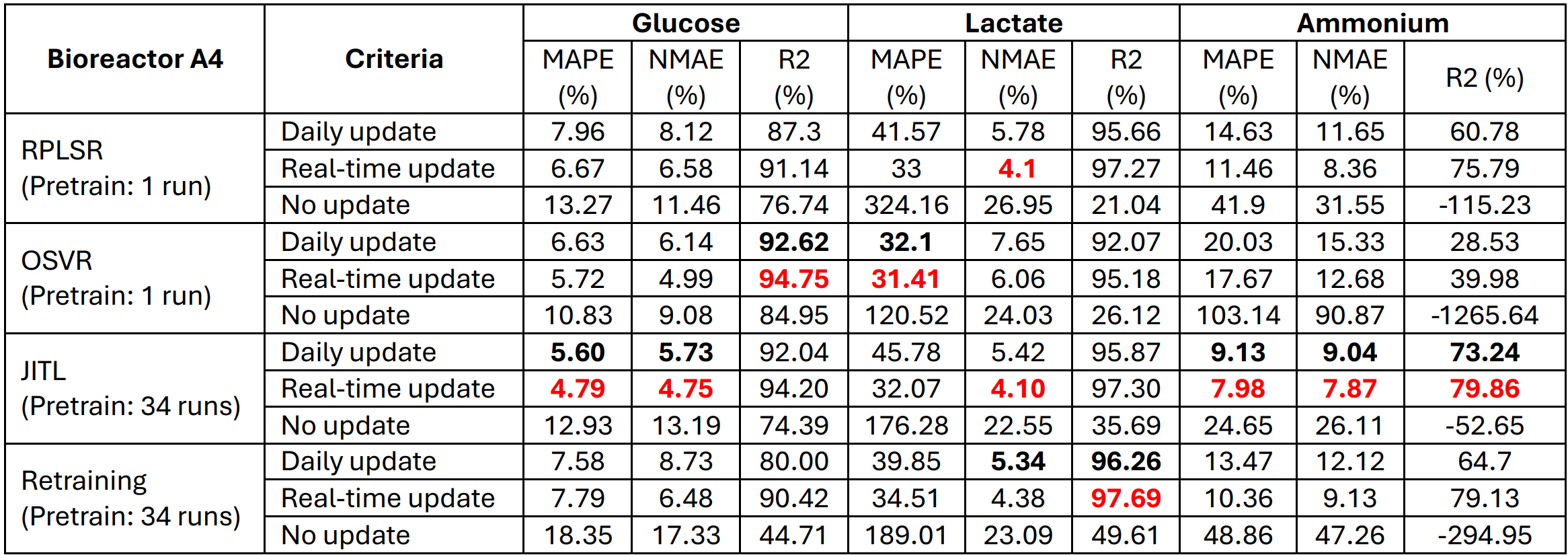}
    \caption{Model Performance on Validation for A3 and A4 Run}
    \label{fig:ValRunMetrics}
\end{table}
Furthermore, models with real-time updates performed the best and consistently outperformed models with daily updates and no updates. These results show that while verifying the validity of data before entering it into the system is important, it is also crucial to minimise the time between data capturing and model updating. Thus, developing automated data validation or outlier detection models would also aid in the development of soft sensors for bioreactors.       

In conclusion, for a new experiment utilising completely new fill/feed media and feeding strategies, JITL and online ML models would be better choices compared to retrained models that combine large historical datasets with new samples. This result on the validation set aligns with the conclusion from \cite{Tulsyan2019}, which suggests that the results from \ref{sec:Training Strategy} show that JITL is unable to beat Global Retraining because the data is relatively homogeneous. The opposite may occur if the data is heterogeneous and contains unseen characteristics. 

\section{Conclusion and Future Work}
\label{sec:conclusion}
\subsection{Lessons Leanred}
This study offered three important lessons on the practical application of machine learning (ML) methods to bioprocess monitoring under conditions of limited data and feedback:
\begin{enumerate}
    \item \textbf{Adaptivity matters more than complexity in low-data regimes:} Just-In-Time Learning (JITL) consistently demonstrated strong performance across different data settings. Its ability to leverage only the most relevant training samples for each prediction made it more resilient to data heterogeneity and infrequent feedback compared to retraining a complex, data-hungry model with all available data. This suggests that, in real-world bioprocess applications where new data may be scarce or delayed, JITL can serve as a robust and lightweight model adaptation method alternative to batch retraining.

    \item \textbf{Data characteristics drive model transferability:} Across all methods, meta-features such as feed composition and process control mode had a notable influence on model generalisability. Models trained under specific process conditions struggled to transfer effectively without incorporating mechanisms (like JITL or online adaptation) to adjust for domain shifts.

    \item \textbf{Hybrid approaches offer promising directions:} Integrating Raman-based predictions with lagged offline measurements improved prediction accuracy across models. This points to the potential of hybrid sensor fusion strategies that combine fast but noisy signals with slower, more accurate ones.
\end{enumerate}

Together, these insights highlight the need for adaptive, context-aware modelling strategies in upstream bioprocess monitoring. They also suggest that method selection should be informed not only by performance metrics but also by operational constraints such as feedback frequency, process variability, and the practical cost of model retraining.

\subsection{Finding and Next Steps}
Overall, this benchmark study has thoroughly benchmarked various dimensionality reduction methods and ML models' effectiveness in building soft sensors for real-time monitoring of bioreactors. The results have shown that soft sensors with reasonable performance can be built using dimensionality reduction methods and ML models with suitable complexity, even in cases where data dimensions exceed the available training samples. At the same time, the meta-analysis on the CSL dataset has shown that the differences in meta-features, such as feed and fill media, could significantly impact the transferability of the models. Similarly, differences in control policy could affect the model's transferability, as shown in meta-analysis results from the IndPenSim dataset. Thus, meta-learning (\cite{lega10,lebu15,albu20}) could be a fruitful direction for the development of bioprocess soft sensors.  

Additionally, the overall results on the CSL dataset reveal that relying solely on Raman data for real-time monitoring is insufficient for building a reliable soft sensor for all offline measurements. At the same time, the results of the AstraZeneca data indicate that a reasonable one-step-ahead prediction can be obtained based solely on lagged observations. Thus, future studies should focus on improved model accuracy by integrating Raman data with other data sources, such as lagged observations. 

Lastly, this study examines the efficacy of handling cold-start scenarios using three training strategies: batch learning, JITL, and OL. Batch Learning is effective with homogeneous data but falters in cold-start scenarios with limited historical data. In contrast, just-in-time learning (JITL) and Online Learning (OL) have demonstrated superior adaptability in these scenarios, with JITL slightly outperforming OL. Given that JITL has shown remarkable adaptability and robustness against cold-start scenarios, another future research direction could be refining its various components. For example, when the hyperparameter $k$ of JITL is set to the total number of data points available, it becomes a batch retraining model. Thus, we can develop a method that dynamically adjusts $k$ based on the homogeneity of the data. JITL may perform as well as, or even outperform, the batch retraining model when the data is homogeneous, while achieving superior performance under a cold-start scenario. That way, JITL becomes a more robust and accurate retraining strategy than all other retraining strategies we explored in this benchmark. Another research direction for JITL is to develop new similarity measures that can produce a unified measurement across heterogeneous data sources. As we discovered from this benchmark study, learning from different data sources that complement each other may help improve the robustness and accuracy of the soft sensor across different offline measurements.

\section*{CRediT authorship contribution statement}
\textbf{Johnny Peng:} Conceptualisation, Methodology, Investigation, Software, Validation, Visualisation, Writing – original draft. \textbf{Thanh Tung Khuat:} Conceptualisation, Methodology, Investigation, Validation, Writing – review \& editing. \textbf{Ellen Otte:} Conceptualisation, Investigation, Validation, Supervision, Writing – review. \textbf{Katarzyna Musial:} Conceptualisation, Investigation, Validation, Supervision, Writing – review \& editing. \textbf{Bogdan Gabrys:} Conceptualisation, Methodology, Investigation, Validation, Project administration, Funding acquisition, Writing – review \& editing. 

\section*{Declaration of Competing Interest}
Ellen Otte is an employee of CSL Innovation Pty Ltd. The other authors declare no competing interests, including no known competing financial interests or personal relationships that could have appeared to influence the work reported in this paper.

\section*{Acknowledgements}
This research was supported under the Australian Research Council's Industrial Transformation Research Program (ITRP) funding scheme (project number IH210100051). The ARC Digital Bioprocess Development Hub is a collaboration between The University of Melbourne, University of Technology Sydney, RMIT University, CSL Innovation Pty Ltd, Cytiva (Global Life Science Solutions Australia Pty Ltd) and Patheon Biologics Australia Pty Ltd.

\bibliography{ref}

\end{document}